\documentclass[12pt,preprint]{aastex}
\usepackage {graphicx}
\usepackage{aastexug}

\shorttitle{Protostellar Jet Model of Chondrule Formation} 
\shortauthors{K. Liffman and M. J. I. Brown}

\def\gapp{\lower 3pt\hbox{${\buildrel > \over \sim}$}}
\def\lapp{\ \lower 3pt\hbox{${\buildrel < \over \sim}$}\ }

\def\beq{\begin{equation}}
\def\eeq{\end{equation}}
\def\beqa{\begin{eqnarray}}
\def\eeqa{\end{eqnarray}}
\def\beqas{\begin{eqnarray*}}
\def\eeqas{\end{eqnarray*}}

\begin{document}

\title{The Protostellar Jet Model of Chondrule Formation}

\author{Kurt Liffman,\altaffilmark{1}and Michael Brown,\altaffilmark{2}}

\altaffiltext{1}{CSIRO/MIT, P.O. Box 56, Highett, Victoria 3190, AUSTRALIA; 
Kurt.Liffman@csiro.au} 

\altaffiltext{2}{Astrophysics Group, School of Physics, University of Melbourne,
Parkville, Victoria 3052, AUSTRALIA}

\affil
\singlespace

\begin{abstract}

A chondrule formation theory is presented where the chondrule
formation zone is located within 0.1 AU of the protosun. This hot, optically thick,
 inner zone of the solar accretion disk is
coincident with the formation region of the protosolar jet. It is suggested,
that chondrules are ablation droplets produced by the interaction
of the jet wind with macroscopic
bodies that stray into the jet formation region.

Provided these droplets are small enough, they will be swept up by the jet wind and subsequently
cool at approximately the same rate as the expanding gas in the jet. There is a
 critical gas density ($\sim 10^{-11}$ g cm$^{-3}$) below which
 chondrules will undergo large, damped oscillations in altitude and thereby suffer reheating.

We claim that it is in the cooler, high altitude regions
above the mid plane of the inner accretion disk that compound chondrules are formed, and the
collisional fragmentation of chondrules take place. Since these processes
 take place in the jet
flow, one can make a prediction for the expected structure of triple compound-chondrules.
For such chondrules, it is suggested that two ``relatively large''
secondary chondrules will avoid
each other. This prediction is valid only if the gas-flow is sufficiently laminar
or if the ``spin-down time'' for a double compound
chondrule is less than the inter-chondrule collision time.

The model assumes that particles, ranging in diameter from 1 $\mu$m to 1 cm,
 can be ejected from the inner-accretion disk by the jet flow, and
that the angular momentum of this material is sufficient to eject it from the jet flow.
Given these assumptions, any material so ejected,
will fly across the face of the accretion disk at speeds greater than
the escape velocity of the system. This material can
only be recaptured  through the action of gas drag.
Such a capture process naturally produces aerodynamic size sorting of chondrules and
chondrule fragments, while
the ejection of refractory dust provides
 a possible explanation for the observed complementarity between matrix and chondrules.

This transfer of material
 will result in the loss of angular momentum
from the upper atmosphere of the outer accretion disk
and thereby facilitate the accretion of matter onto the protosun.
 
\end{abstract}

\newpage

{\bf PREFACE}

In this paper we discuss a relatively new chondrule formation model, which
uses a hypothetical protosolar jet as the main chondrule formation mechanism.
As this model is in the earliest stages of development, this particular
monograph should not be treated as the forever, enduring, last word on the
subject. Instead, we hope that our discussion will demonstrate the
explanative and predictive power of this model, and possibly prompt
our colleagues to view chondrule formation from a different
point of view as compared to other theories.

    Some readers would argue that fewer, not more, chondrule formation
theories are required. Certainly with nearly twenty different published
theories, how can one determine which theory, if any, is the correct one?
The answer, of course, is to embrace the theory that can not only explain
many of the observed data, but also predict new facts. So in this study,
we will attempt to explain:
\par
(1) the chondrule formation process;
\par
(2) the chondrule cooling rate;
\par
\hangindent = 0.5 in
\hangafter = 1
(3) the collisional fragmentation of chondrules and
the formation of adhering compound chondrules, where the secondary (and
smaller) chondrule was plastic at the time of collision;
\par
(4) chondrule reheating;
\par
(5) the chondrule upper size limit;
\par
(6) chondrule size sorting,
\par
\hangindent = 0.3 in
\hangafter = 0
and

(7) the complementary chemical composition of matrix and chondrules.
\par
We will also discuss a possible link between protostellar jets and disk accretion,
where a protostellar jet may inject material with low angular momentum into
the upper atmosphere of an accretion disk, thereby enhancing stratified accretional flow,
i.e., the upper layers of the disk accrete onto the protostar more readily than do
the layers adjacent to the mid plane of the accretion disk.

As for predictions, we will describe the expected structure of triple
compound chondrules, i.e., where two small secondaries are adhering
to a large primary. We suggest that, for such a compound chondrule, the two
secondary chondrules will tend to ``avoid'' each other. We give the mathematical
expression for the minimum avoidance angle between the two secondaries and
discuss under what circumstances this prediction will break down. We also show
that the experimental data collected so far (Unfortunately, for only
four such chondrules) have
structures that are consistent with the predicted structure. It is our fond hope,
that we will inspire some of our colleagues to collect much more data, so
that the validity or otherwise of this prediction can be confidently determined.

\section{JET ABLATION
\label{sec:jet_ablation}}

In the late 1970's, it was found that
protostellar systems formed bipolar outflows of material,
i.e., the gas flows travelling in two opposing directions, approximately perpendicular to
the plane of the accretion disk (for a review, see Beckwith and Sargent 1993).
These flows were detected via the CO rotation lines and consisted mostly
of molecular material. The flow speeds of the bipolar outflows were found
to be around 20 km s$^{-1}$, and by dividing their length by the observed
flow speed, one could deduce a ``dynamic'' lifetime for these flows of
around 10$^5$ years. Later, unbiased surveys of protostellar systems
suggests that \it all \rm protostellar systems undergo some form of
bipolar outflow stage (Fukui \it et al. \rm 1993).

Often, but not always, one can find within a bipolar outflow a
 faster, more collimated flow known as a protostellar jet. Protostellar
jets are usually detected in the SII, OI and H$\alpha$ lines and are
consequently referred to as optical jets. They have observed
wind speeds in the range of 100 - 400 km s$^{-1}$, they eject a
large amount of gas (total mass $10^{-3} - 10^{-1}$ M$_\odot$),
are quite energetic (total kinetic energy 10$^{44} - 10^{46}$ erg),
long-lived phenomena (10$^6 -10^7$ years) that
exist at the very earliest stages of star formation
(Cabrit et al. 1990, Edwards, Ray and Mundt 1993). They occur not only
in the massively active FU-Ori stages of star formation, but they also
are to be found in the more quiescent, Classical T-Tauri Star
(CTTS) stages. They appear to
be generated by the interaction between protostars and accretion disks,
within 10 stellar radii ($R_*$) of the protostar (Hartmann 1992).
\par
What is the connection between bipolar outflows and protostellar jets?
There is a growing consensus that bipolar outflows are a byproduct of
protostellar jets (Snell \it et al. \rm 1980, Shu \it et al. \rm 1993),
where the protostellar jet sweeps up ambient molecular cloud material
into two thin shells, which manifest themselves as the observed bipolar
lobes of CO emission. Once the molecular cloud material has been swept
away (on a timescale of 10$^5$ years),  the bipolar outflow disappears,
 leaving the protostellar jet to erratically fire away for a further
10$^6$ - 10$^7$  years.
\par
Our interest in protostellar jets is sparked by the following mass
processing argument. Suppose the solar nebula formed
a protosolar jet. Such a jet would have ejected $10^{-5} - 10^{-3}
$ M$_\odot$ of ``rocky'' material (i.e. all elements excluding H and He)
from the solar nebula.
If only 10\% of this material were to fall back to the solar nebula
then we would have  $10^{-6} - 10^{-4}$ M$_\odot$ of rocky ( possibly
refractory) material being contributed to the solar nebula over a
10$^6 -10^7$ year period. Given that the ``rock'' mass of the planets is of order
$10^{-4}$ M$_\odot$, a protosolar jet may have made a significant
contribution to the chemical structure of the solar nebula. Indeed,
it is possible that ejecta from the jet flow may have been incorporated into
the best preserved samples of the solar nebula: the chondritic
meteorites.

Of course, such an argument does not prove that protostellar jets formed chondrules,
but only provides a plausible basis for constructing a theory of
chondrule formation. To create such a theory,
 our first task is to investigate the thermal environment
of the jet formation region.
This is a function of the formation distance of a protosolar jet
 from the protosun. A distance which is uncertain, but
model fits to the Li 6707 \AA, Fe 1 4957 \AA, Fe II 5018 \AA $\ $ lines in FU Ori,
plus the lack of
extinction and infrared excess in the wind, suggests a formation distance of
$\leq$ 10$R_*$ ($\approx$ 0.1 AU for the solar nebula), where $R_*$ is the
radius of the protostar (Hartmann 1992). The temperature of the disk surface ($T_s$)
at such distances, is given (approximately) by the formula (Frank, King and
Raine 1992)

\beq
T_s(r) = 850  \ {\rm K } \ \left( \frac{ \left( M/{\rm M}_\odot \right)
\left(\dot M/10^{-7} {\rm M}_\odot \ {\rm yr}^{-1} \right)}
{\left(r/0.1 \hbox{ AU} \right)^3 }
\right)^{1/4}
\left( 1 - (R_*/r)^{1/2} \right)^{1/4} \ ,
\label{eq:1.1} 
\eeq

where $M$ is the mass of the protostar, $\dot M$ is the mass accretion rate,
and $r$ is the radial distance from the protostar's centre. The maximum temperature
of the midplane of the accretion disk ($T_m$) can be approximately
determined from the formula
\beq
 T_m \approx T_s (\eta \tau)^{1/4}  \ , 
\label{eq:1.2}
\eeq
where $\eta$ is some number of order 1, and $\tau$ is the optical depth
(\it ibid.\rm, and Cassen 1993). Since $\tau$ may have values as high as
$10^4$, it is probable that
 at distances of order 0.05 AU
from the protostar, the temperature at the midplane maybe high enough to melt or
vaporise any rocky bodies that happen to be in that section of the accretion disk.
This raises the possibility that the protostellar jet winds may ablate droplets
from the surface of rather warm ``rock'' bodies and if the droplets are small
enough, they may
be ejected from the inner accretion disk by the drag force of the jet flow.

It may be asked, however, why embrace the idea that chondrules are ablation
droplets?
We are encouraged to advance such a hypothesis, because
 chondrules appear to have been extruded or drawn from
one or more extended magma bodies (Dodd and Teleky 1967).
Also, an ablative process readily forms chondrule-like spheres from meteors
(Brownlee \it et al. \rm 1983). These ``ablation-spheres'' are chondrule-like in size
and shape. They also share some physical similarities, e.g., they often contain relict
grains. Despite this supportive evidence,
there are, however, some significant problems to
overcome before one can believe the ablation hypothesis.

First, we require the appropriately sized rocky-bodies to enter into the
jet formation region.
Protostellar systems generally have accretion disks
and large bodies could be caught up in
the inward accretional flow. However, the existence of chondrules in single ordinary
chondrites that have (admittedly uncertain) $^{129}$Xe$^*$/$^{127}$Xe ages
ranging over 10$^7$ years (Swindle \it et al. \rm 1991, Swindle and Podosek 1988),
suggests that the accretional flow in the solar nebula was small enough to allow the
survival
of the chondrite precursors. Indeed, it is possible that such accretional inflow may
have been dependent on
height within the accretion disk, where only the
gas at higher altitudes took part in the accretional flow (see \S 5).

Another mechanism for radial migration is that of gas drag due to the velocity
difference between nebula gas and the near Keplerian motion of large bodies
around the protosun.
The infall velocity for material subject to this gas drag is ( Whipple 1973,
Weidenschilling 1977)
\beq
\frac{dr}{dt} \approx - \frac{2r\Delta V}{ V_{Kep} t_D} \ , 
\label{eq:1.3}
\eeq
where $V_{Kep}$ is the Keplerian velocity, $\Delta V $ is ($\approx
10^{-3}V_{Kep} $) the difference between the gas angular velocity and $V_{Kep}$,
while $t_D$ is
the time scale for gas drag to influence the
motion of the body. In the Stokes drag regime,
\beq
 t_D = \frac{ 2r_{fod}^2\rho_b}{ 9\eta_{g}} \ , 
\label{eq:1.4}
\eeq
where $\rho_b$ is the mass density of the body and $\eta_{g}$ is the viscosity
of the gas, and $r_{fod}$ is the radius of the ``chondrule-fodder'' body.
 Using the appropriate formula for the gas viscosity
(e.g. Eq. (2.3.5) of Liffman 1992),
one can obtain, from Eq.~(\ref{eq:1.3}),  the timescale for orbital decay ($\tau_{decay}$)
\beq
\tau_{decay} \approx 10^5 
\left(\frac{ \left( r_{fod}/ 10^3 \hbox{ cm} \right)
\left(\rho_{fod}/1 \hbox{ g cm}^{-3} \right)
}{
\left((T_g/100)^{1/2} \hbox{ K} \right)} \right) \hbox{  yrs} \ , 
\label{eq:1.5}
\eeq
with $T_g$ being the temperature of the gas.

So, objects with a radius of around 10 m would fall into
the Sun on a timescale of about 10$^5$ years. This analysis, however, ignores
the change in the size of the body due to accretion of dust from the solar nebula.
If the dust to gas mass ratio is of $\sim$ 1,
accretion of material will increase the size of the body and eventually
stop its infall into the Sun (Weidenschilling 1988). On the other hand fragmentation
due to inter-body collisions would have been a source of smaller material which may have
eventually reached the boundary layer between the protosun and the solar nebula.

There may have been other mechanisms that
brought macroscopic material
into the inner regions of the solar nebula. We will simply note that it is
 a plausible assumption and one we require for our model.

Once we have these large bodies in the jet-formation region, we are faced
with the following scenario: the high speed jet-flow will probably occur at $z \ge$
the scale height of the accretion disk, $H$,
since the outflow jet must be governed by the conservation of mass equation,
and so high velocities will only occur when $\rho_{g}(z) \ll \rho_{g}(0)$.
On the other hand, our ``chondrule-fodder'' bodies will be
located on or near the midplane of
the nebula ($z = 0$). This separation between the critical points in the
wind and the chondrule-fodder bodies
would appear to be the death-knell for our ablation hypothesis - for how else
can one produce ablation droplets if the hypothetical windflow is nowhere
near our chondrule-producing planetesimals?

To answer this, we examine protostellar jet theory.

\begin{figure}
\epsscale{0.4}
\begin{center}
\plotone{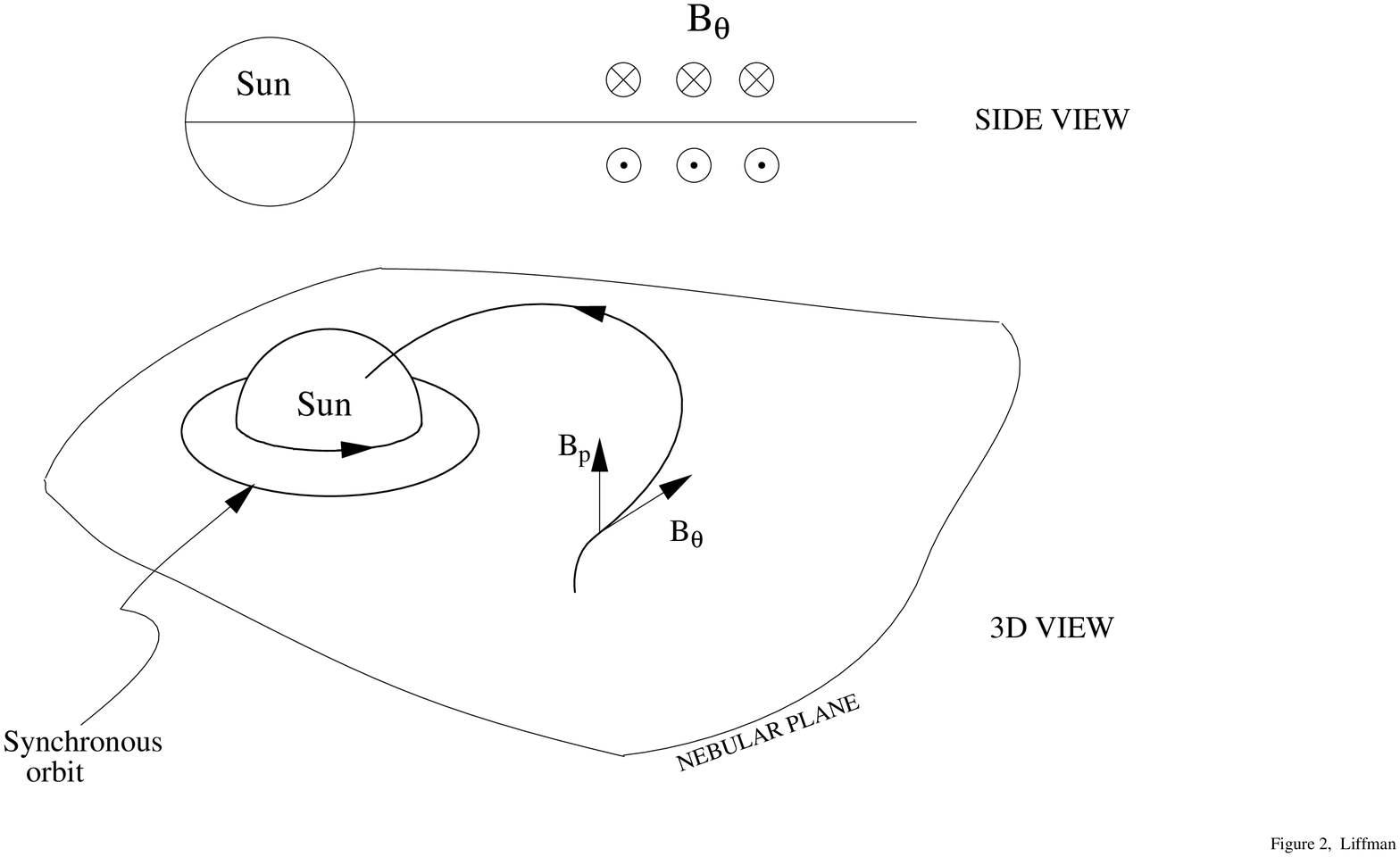}
\end{center}
\caption{
The dipole field of the protosun interacts with the 
accretion disk, such that for
distances greater than a critical distance, ($\sim$ 0.04 AU) from the protosun,
the poloidal field of the Sun is ``wrapped-up'' by the accretion disk.%
This process may cause the central plane of the inner accretion disk to become
a ``current sheet'' produced by the merging magnetic field lines.
This merging will be due to magnetic diffusion and the switch in the sign of the toroidal
 field as one traverses the central plane.} 
\label{fig:proto_sun}
\end{figure}

Most protostellar outflow
models assume that magnetic fields provide the main coupling mechanism
between the accretion disk and the outflow ( for recent reviews, see
Bicknell 1992, K$\ddot{\hbox{o}}$nigl and Ruden 1993 plus
Shu \it et al. \rm 1993).
In the most recent models, the magnetic driving
 mechanism involves the interaction between the dipole field of
the protosun and the inner accretion disk (e.g. see Lovelace \it et. al. \rm 1991,
and Shu \it et. al. \rm 1994). Studies of this protosun-nebula
interaction can be traced back
to the work of Freeman (1977),
who assumed that the dipole field
of the protosun was able to thread the inner (partially ionized) disk as shown in Fig.~\ref{fig:proto_sun}.
In effect, the magnetic field of the protosun is ``tied'' to its surrounding
accretion disk.

In Freeman's model, the dipole field of the protosun rotates
 (approximately) with a rigid body velocity,
so there exists a point away from the protosun ($\sim$ 0.04 AU), where the speed of the field
sweeping over the disk
equals the (approximately) Keplerian velocity of the disk.
In Fig.~\ref{fig:proto_sun}, 
this position is denoted as the ``synchronous orbit''.
For distances $r$ greater than this synchronous distance, the magnetic
field of the protosun has a greater angular velocity than the
the accretion disk and so the magnetic field becomes ``wrapped up'',
\it i.e, \rm the purely poloidal magnetic
field of the protosun is converted into a toroidal field in the disk.

In Fig.~\ref{fig:proto_sun} ,
 we show a side view of the resulting field structure. The toroidal field
 becomes the dominant field in the disk, and its direction reverses
 as one traverses the central plane. For such a configuration,
the central plane of the nebula can become a ``current sheet'' where magnetic fields
reconnect and the magnetic energy so released is converted into particle  energy.

As discussed in Priest (1994) the particle flow velocity ($v_f$)
obtained from the merging of magnetic fields ($B$) is close to the characteristic
speed of a magnetic medium (\it i.e., \rm the Alfv\'en speed $C_A$), which in
mathematical notation has the form
\beq
v_f \approx C_A = \frac{B }{ \sqrt{\mu_0 \rho_g }} =
3 \ \left( \frac{B }{ 100 \ \hbox{G}} \right)
\left(\frac{10^{-8} \ \hbox{g cm}^{-3}}{ \rho_g }\right)^{1/2} \ \hbox{km s}^{-
1} \ , 
\label{eq:1.6}
\eeq
where $\rho_g$ is the mass density of the gas and $\mu_0$ is the
permeability of free space.
In the accretion disk, the direction of the reconnection flow would be
roughly parallel to the central plane,
with a small
component in the $z$ direction.

To obtain ablative behaviour from gas flows with
gas densities in the range of $10^{-6}$ to $10^{-8}$ g cm$^{-3}$,
the required gas speeds range from 1 to 20 km s$^{-1}$ (Liffman 1992).
Comparing this to Eq. (\ref{eq:1.6}) suggests that these ``reconnection'' flows may
have the required densities and speeds to ablate the planetesimals. Some of the
ablation droplets so produced would then, presumably, be caught up in the flow
field that eventually becomes the main protostellar jet flow.

It may turn out, however, that outflows
are not powered by ``wrapped-up'' toroidal fields. Our
purpose in describing this particular model is to demonstrate that, in the outflow region,
the central plane of the accretion disk may be a highly active region, where conditions
are conducive to chondrule formation by ablation.

\section{CHONDRULE COOLING}

A long standing problem in chondrule formation is how to
explain the slow rate of chondrule cooling. Experimental simulation
of chondrule formation suggests that chondrules cooled at a rate of
 5 to 2000 $^o$C/hour (Hewins 1988). Such
cooling rates are 3 to 5 orders of magnitude smaller than those expected for an
isolated black body radiating heat directly into space. It would appear
 that chondrules were formed in a hot optically-thick medium or were
produced in close proximity to other chondrules so that mutual radiation
could damp the cooling rate.

In our model, chondrules are produced in the hot, optically-thick midplane of
the inner accretion disk around a protostar. They are produced by the
ablative interaction between the initial stage of a
 protostellar jet wind and rocky chondrule ``fodder'' bodies that happen
to stray into the jet formation region. If the subsequent droplets are
small enough, they will be swept up by the wind and begin to move with
the jet flow.

As the jet flow moves away from the midplane of the
accretion disk, it is likely that the gas in the flow will cool,
since work is being done to expand the
gas. Consequently, if the molten droplets are close to thermodynamic equilibrium
with the surrounding gas flow, they must also cool at just about the same rate
as the expanding gas flow.

Naturally, to model this process we require a model of a protostellar jet.
Unfortunately, a comprehensive
theory of protostellar jets is unavailable at this time.
To partially circumvent this difficulty, we adopt a simple parameterised model of
an outflow. First, we describe our system with cylindrical coordinates, the plane
$z = 0$ being the midplane of the accretion disk with the protostar residing
at the origin (see Fig.~\ref{fig:full_scheme}).
With this coordinate system, the steady state form of the continuity or
mass conservation equation
($\nabla \bullet (\rho_g \hbox{\bf v$_g$}) = 0$)
is easily solved, for an axisymmetric flow, to give the equation

\beq
 2 \bar{\rho}_g(z) \bar{v}_{gz}(z) \pi r(z)^2 = 
\hbox{constant} = \dot{M}_o \ , 
\label{eq:2.1}
\eeq

where $\bar{\rho}_g(z)$ is the ``$r$-averaged'' gas mass-density as a function of $z$,
 $r(z)$ is the cylindrical radius of the outflow,
$\bar{v}_{gz}(z)$ is the $z$ component of the ``$r$-averaged'' gas velocity,
 and
$\dot M_{o}$ is the mass-loss rate of the outflow. The factor of two in Eq.~(\ref{eq:2.1})
arises, because the protostellar jet is produced from both sides of the accretion
disk.
\par
To model the gas density and gas velocity profile within the disk, we assume that the disk
constrains the radial size of the protostellar jet. If we now let
\beq
 \bar{\rho}_g(z) \approx \bar{\rho}_g(z_1)\left({z_1 \over z} \right)^m
\ ,\qquad z \ \lapp \ z_1 \ , \quad m > 0 \ ,
\label{eq:2.2}
\eeq
where $z_1$ is the distance above the disk midplane where we can best define the
flow variables.
We can substitute Eq.(\ref{eq:2.2}) into Eq.(\ref{eq:2.1}) and obtain
\beq
\bar{v}_{gz}(z) \approx \bar{v}_{gz}(z_1)\left(\frac{z}{z_1} \right)^m
\ ,\qquad z \ \lapp \ z_1 \ , \quad m > 0 \ .
\label{eq:2.3}
\eeq
These solutions break down as ${z \to 0}$, this being the price we pay for not
solving the momentum and energy equations. Equations (\ref{eq:2.2}) and (\ref{eq:2.3}) 
are only a
first order fit to the mass conservation equations. However, they do,
 approximately,  satisfy the
density-velocity  relationship observed in thermally driven winds, i.e.,
as $\rho_g$ decreases, $v_{gz}$ increases.
\par
If the gas pressure ($P_g$) can be globally modelled as a polytropic gas, then
\beq
 P_g = \kappa \rho_g^\gamma \ , 
\label{eq:2.4}
\eeq
where $\kappa$ and $\gamma$ are constants ($\gamma$ is the ratio of the
specific heats if the jet is adiabatic). Using the ideal gas law with Eqns (\ref{eq:2.2}) 
and (\ref{eq:2.2}),
we can obtain an expression for the gas (and chondrule) temperature
 ($T_g$) as a function of $z$.
\beq
 T_g(z) \approx T_g(z_1)\Bigl({z_1 \over z}\Bigr)^{m(\gamma - 1)} \ . 
\label{eq:2.5}
\eeq
The gas flow and entrained chondrules decrease in temperature as they travel
 away from the central plane of the accretion disk.

If we assume that the particles are entrained with the flow, and they start from $z = 0$, then
we can use Eqns (\ref{eq:2.3}) and (\ref{eq:2.5})
to give the gas and chondrule temperature as a function of time ($t$). Solving for
$t$ in Eq.~\ref{eq:2.3} gives
\beq
 \frac{z }{ z_1} = \left[ \frac{t }{ \tau_1} \right]^{1/(1-m)}  \qquad 0<m<1 \ , 
\label{eq:2.6}
\eeq
where $\tau_1 = z_1/((1 - m) \bar{v}_{gz}(z_1))$. Substituting the above equation into 
Eq.(\ref{eq:2.5})
gives,
\beq
T_g(t) = T_g(z_1)\left[ { \tau_1 \over t} \right]^{ {m(\gamma-1) \over (1-m)}  } 
\qquad 0<m<1. 
\label{eq:2.7}
\eeq
The gas and chondrule cooling timescale is determined by the value of $\tau_1$ and
$ T_g(z_1)$. As we shall show in the next section, for $ T_g(z_1) \sim 1200$ K, then
$z_1 \sim 10^4-10^5$ km and $\bar{v}_{gz}(z_1)) \sim 0.1 - 1$ km s$^{-1}$. As a
consequence, a
minimum parameterization for $\tau_1$ is
\beq
 \tau_1 = 10^4 {\rm s} \ {(z_1/10^4 \ {\rm km}) \over (1-m)(\bar{v}_{gz}(z_1)/ 1
\ {\rm km}{\ \rm s}^{-1}) }. 
\label{eq:2.8}
\eeq

The general decrease in gas temperature, and chondrule temperature, as a function
of $z$ within or near the accretion disk has important consequences for the
collisional interactions between chondrules. Indeed, a simple analysis of
this phenomenon leads to a model which can possibly explain the observed
structure of compound chondrules and leads to a prediction for the physical
structure of triple compound chondrules.

\section{CHONDRULE COLLISIONS}

\subsection{Hover particles}

Let us consider a chondrule that has just been ablated from its parent body.
We assume, that the chondrule will be accelerated by the gas jet so that its
motion is, initially, in the $z$ direction.
The equation of motion for the particle,  parallel to the $z$ axis,  is given by
\beq
 m_p\ddot{z} = \frac{C_D}{ 2} \rho_{g} 
(v_{gz}- \dot{z})^{2} \pi a_{p}^{2}
- \frac{ GMm_pz }{ (z^{2}+r^{2})^{3/2}} \ , 
\label{eq:3.1.1}
\eeq
where $C_D$ is the coefficient of gas drag, while $m_p$ and $a_p$ are the mass and
radius of the particle, respectively.
\par
Suppose the particle reaches a state where the gas drag is balanced by
gravity, so that $\ddot{z}=0$, and $\dot{z}=0$.
  Our equation of motion becomes (for $z\ll r$)
\beq
0 \approx \frac{C_{D} }{ 2} \rho_{g} v_{gz}^{2} \pi a_{p}^{2}
- \frac{GMmz_{h} }{ r^3} \ , 
\label{eq:3.1.2}
\eeq
where $z_{h}$ is the value of $z$ for the hovering particle.
For $z$ close to the midplane, we should expect that $v_{gz}$ will
be less than the sound speed, so if the mean free path of the gas ($l$)
satisfies the relation
\beq 
l/2a_p \ \gapp 10 \ 
\label{eq:3.1.3}
\eeq
then $C_D$ has the Epstein (1924) form
\beq
 C_D \approx \frac{8 }{ 3v_{gz}} \sqrt{\frac{8kT_g }{ \pi m_g} } \ ,
 \label{eq:3.1.4}
\eeq
where $k$ is the Boltzmann constant and $m_g$ is the mass of a gas particle
(in this case we assume monatomic hydrogen). Using Eqns (\ref{eq:3.1.2}), 
(\ref{eq:3.1.4}) and (\ref{eq:2.1})
we find
\beq 
z_{h} \approx \left(\frac{8 kT_{g} }{ \pi m_{g}}\right)^{1/2}
\frac{\dot{M}r }{ 2\pi GM\rho_{p} a_{p}} \  
\label{eq:3.1.5} 
\eeq
or
\beq
z_{h} \approx 5 \times 10^{4}
\frac{(\dot{M}/10^{-8}{\ \rm M}_\odot/{\ \rm yr})(r/0.1{\ \rm AU})(T_g/10^{3}{\ \rm K})^{1/2}
}{
(M/{\rm M}_\odot)(\rho_{p}/1 {\ \rm gm} {\ \rm cm}^{-3})(a_{p}/0.1 {\ \rm cm} ) } {\ \rm km} \ .
\label{eq:3.1.6}
\eeq
Thus, $z_h \sim 10^{-4}$ AU $\ll  0.1$ AU $\sim r$ as is
 required from the assumption leading to
 Eq. (\ref{eq:3.1.2}). Now, $z_h$ is inversely proportional to the radius of
 the particle ($a_p$), so smaller
 particles will have higher hover heights than larger particles of the same density.
 To reach their respective
 hover altitudes, the particles will move relative to each other and may undergo
 collision interaction. If the collisional velocities are small enough and the temperatures
high enough, we may have chondrules fusing together to form compound chondrules.

\subsection{ Compound Chondrule Formation}

Studies of compound chondrules have classified them into two general
 types:
\it enveloping \rm, where one chondrule envelopes the other and \it adhering \rm,
where one chondrule forms a ``bump'' on the other (Wasson 1993). In this paper,
we restrict our attention to adhering compound chondrules.
\par
 Most adhering compound chondrules consist of a smaller chondrule stuck to a larger particle,
where the smaller chondrule was plastic at the time of collision (Wasson \it et al. \rm 1995).
This fact immediately poses a major problem for most chondrule formation theories which use
simple ``flash'' heating scenarios. Since, if the chondrules were all formed in
the one flash heating event, then we should expect the larger chondrules to be plastic
at the time of collision.

In the Jet model, chondrules are continuously formed by an ablative process as
chondrule fodder bodies stray into the jet formation region. As discussed in
the derivation of Eq. (\ref{eq:3.1.6}), it is reasonable to assume that 0.1 cm
particles will be supported by the jet flow at a $z$ distance of 10$^{-4}$ to 10$^{-3}$ AU
above the midplane of the accretion disk. Let us suppose that at such distances,
the temperature of the gas in the Jet stream is around 1200 K - the approximate
solidus temperature of a chondrule.
\par
Assuming a fairly steady jet flow, a hover particle in such a position should have
sufficient time to equilibrate with the gas temperature and be relatively non-plastic.
The hover particle will only see smaller particles flying past it in the
jet flow, since larger particles of approximately the same mass density will have
a lower hover altitude.

\begin{figure}
\epsscale{0.3}
\begin{center}
\plotone{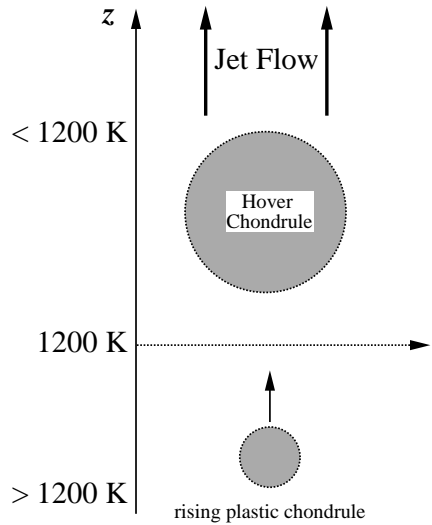}
\end{center}
\caption{
 Schematic depiction of compound chondrule formation in a jet flow.
The gas flow is depicted within the accretion disk, where the gas temperature
decreases as $z$ (the distance from the midplane of the disk) increases.
A ``large'' hovering chondrule, with a temperature near or at the solidus temperature of
1200 K, collides with smaller chondrules that are entrained
in the gas flow. Due to thermal inertia, the smaller chondrules are warmer
 and more plastic than the hover chondrule. In this way, one can obtain the observed
phenomena that smaller chondrules were plastic at the time the compound chondrule was formed.
}
\label{fig:chondrule_collision}
\end{figure}
As these smaller particles fly past the hover particle, the thermal inertia
of these particles
 will give them higher temperatures, and therefore greater plasticity than the hover particle.
If the hover particle collides and fuses with these particles, then the observed
structure of adhering compound chondrules will be obtained (for a schematic
depiction of this process, see Fig.~\ref{fig:chondrule_collision} ).

To give this idea a quantitative context, we note that
the equation for the rate of temperature change of a
spherical chondrule has the form
\beq
 \frac{4 }{ 3}\pi a_p^{3}\rho _{p}C_{V} \frac{dT_{S} }{ dt}
= \Lambda 4 \pi a_p^{2} q
+ 4 \pi a_p^{2}\epsilon_a \sigma T_{e}^{4} -
4 \pi a_p^{2}\epsilon_e \sigma T_{S}^{4} \ ,
\label{eq:3.2.1}
\eeq
(Liffman 1992 and references therein).

The left hand side of the equation describes the time rate of change
of the heat energy of the body, where $a_p$ is the radius of the particle,
$\rho_{p}$ is the density of the particle, $C_{V}$ is the specific heat per
mass of the body ($\sim$$10^{7}$ erg g$^{-1}$ K$^{-1}$ for chondrules), $T_{S}$ is
the surface temperature of the body, and $t$ the time. The first term on the
right hand side describes the energy added to the body by the gas/body
interaction, where $\Lambda$ is the heat transfer coefficient and
$q$ is the gas/body heat transfer rate per unit surface area.
The last two terms are radiation terms, where
$T_{e}$ is the radiation temperature of the surrounding environment,
$\epsilon_{a/e}$ is the absorptivity/emissivity of the surface
(absorptivity and emissivity are assumed to be the same in this case), and
$\sigma$ is the Stefan-Boltzmann constant. In the following discussion,
we will assume local thermodynamic equilibrium applies and that $T_e = T_g$.

For chondrule formation, the size of the chondrules ($<1$ cm)
and the expected low gas density ($<10^{-6}$ g/cc)
indicate that the mean-free path of the gas is large relative to the size
of the chondrules.
For this ``free molecular'' flow regime
 $\Lambda \approx 1$ and $q$ has the form
\beq
q = \rho _{g} \vert v_{g} - v_{p} \vert (T_{rec} - T_{s}) C_{H} \ , 
\label{eq:3.2.2} 
\eeq
(Hayes and Probstein 1959, Probstein 1968)
where $\rho_{g}$ is the mass density of the gas, $ v_{g} $ is the
gas velocity, $ v_{p} $ is the particle velocity, $ T_{rec} $ is the
recovery temperature, and $C_{H}$ is the heat transfer function
for free molecular flow. $T_{rec}$ and $C_{H}$ have the forms
\beq
T_{rec} = \frac{T_{g} }{ \gamma +1}
\left[ 2\gamma + 2(\gamma -1)s_{gp}^{2}
- \frac{\gamma -1 }{
0.5 + s_{gp}^{-2} + s_{gp} \pi ^{-0.5} \exp(-s_{gp}^{2})/{\rm\,erf}(s_{gp})}
\right], 
\label{eq:3.2.3}
\eeq
and
\beq
C_{H} = \left( \frac{\gamma +1 }{ \gamma -1} \right) 
\frac{k }{ 8m_{g}s^{2}_{gp}}
[\pi^{-0.5} s_{gp} \exp(-s^{2}_{gp}) + (0.5 + s^{2}_{gp}) {\rm\,erf}(s_{gp}) ],
\label{eq:3.2.4}
\eeq
(\it ibid. \rm)
where $T_{g}$ is the gas temperature, $m_{g}$ the mean gas particle mass
($ \sim 1.66 \times 10^{-24}$g), $\gamma$ is the ratio of specific heats, $k$
is Boltzmann's constant, erf$(s)$ is the error function
$ (= {2 \over \sqrt{\pi}} \int_0^sexp(-t^2)dt$ ), and
$s_{gp}$ is the ratio of the relative streaming gas speed and the most probable
Maxwellian gas speed,
\beq
 s_{gp} = \frac{\vert v_{g} - v_{p}\vert }{ \sqrt{2kT_{g}/m_{g}}} . 
\label{eq:3.2.5}
\eeq
We are considering flows at the very base of the protostellar
jet, so we should expect that $s_{gp} \ll 1$. Thus, $q$ becomes
\beq
 q \approx \rho_g(T_g - T_s)\Bigl({\gamma + 1 \over \gamma - 1} \Bigr)
 \sqrt{\frac{k^3 T_g }{ 8 \pi m_g^3}}
 \ . 
\label{eq:3.2.6} 
\eeq
or
\beq
 q \approx 5 \times 10^7 \ \Bigl(\frac{\rho_g }{ 10^{-8} {\ \rm g cm}^{-3}} \Bigr)
\biggl( \Bigl( \frac{T_g }{ 10^3 {\ \rm K} } \Bigr) -
\Bigl( \frac{ T_s }{ 10^3 {\ \rm K} } \Bigr) \biggr) 
\Bigl( \frac{\gamma + 1 }{ \gamma - 1} \Bigr)
\sqrt{ \frac{ T_g }{ 10^3 {\ \rm K} } } \ {\rm erg s}^{-1} \ {\rm cm}^{-2}
\ {\rm s}^{-1}. 
\label{eq:3.2.7}
\eeq

When $\rho_g \ \gapp \ 10^{-7}$ g cm$^{-3}$, $\gamma$ = 5/3 (monatomic gas),
 and $T_g \approx 1200$ K, then
the $q$ term in Eq. (\ref{eq:3.2.1}) dominates the radiative terms, since
\beq
\sigma T_g^4 \approx 5.7 \times 10^7 \Bigl( \frac{ T_g }{ 10^3 \ {\rm K}}
\Bigr)^4 \  {\rm erg s}^{-1} \ {\rm cm}^{-2}
\ {\rm s}^{-1}. 
\label{eq:3.2.8}
\eeq

So Eq.(\ref{eq:3.2.1}) becomes a first-order linear differential equation in $T_s$,
and has the solution
\beq
T_s(t) \approx T_g + (T_s(0) - T_g) { \rm exp}(-t/\tau) \ , 
\label{eq:3.2.9}
\eeq
where $\tau$ has the form
\beq
 \tau = \frac{a_p \rho_p C_V }{ 3 \Lambda \rho_g} \Bigl(
\frac{\gamma - 1 }{ \gamma + 1} \Bigr)
\sqrt{ \frac{8 \pi m_g^3 }{ k^3 T_g} } \ . 
\label{eq:3.2.9}
\eeq
or
\beq
\tau \approx 5 \ \frac{ (a_p/0.1 {\rm \ cm})(\rho_p/3 {\rm \ g \ cm}^{-3})
(C_V/10^7 {\ \rm erg \ g}^{-1}{\rm \  K}^{-1}) }{
(\rho_g/10^{-8} {\ \rm g cm}^{-3}) } \Bigl( {T_g \over 10^3 {\rm \ K} } \Bigr)
\ s 
\label{eq:3.2.10}
\eeq

\begin{figure}
\epsscale{0.4}
\begin{center}
\plotone{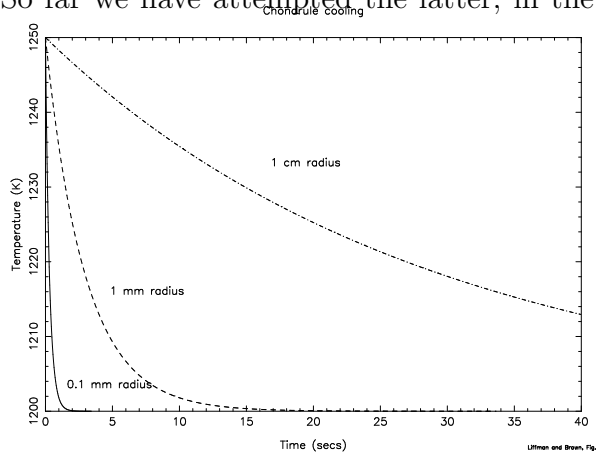}
\end{center}
\caption{
 The radiative cooling of 3 particles of different radii from 1250 K to
1200 K.
}
\label{fig:chondrule_cool}
\end{figure}

So, if we have $\rho_g \  \gapp \ 10^{-7}$ g cm$^{-3}$ then particles with a radius
 between 0.01 cm and 0.1 cm will have a minimum temperature equilibration timescale
 of 0.5 to 5 s. Similar timescales are obtained for the low density case
($\rho_g \ll 10^{-7}$ g cm$^{-3}$), where the radiative terms in Eq.(\ref{eq:3.2.1})
dominate the ``$q$'' convective term. In this case, Eq.(\ref{eq:3.2.1})
 can only be solved
numerically. The results from this calculation are shown in 
Fig.~\ref{fig:chondrule_cool}.

As can be seen, The e-folding
timescales in the radiative case turn out to be about the same as those for the
pure conduction case.
One should note that these are minimum timescales, because we have neglected
the latent heat of fusion and the
diffusion time for heat to travel from the surface of the particle to its
centre.

Despite these caveats,
 it is probable that the timescales for thermal equilibration are not much
greater than as stated in Eq.(\ref{eq:3.2.10}). In such circumstances, we can only expect the
smaller chondrule to be plastic at the time of collision
if, in the collision zone, there is a sharp decline in gas temperature.
 We suggest that a temperature drop of 10 to 100 K s$^{-1}$ is required for
compound chondrules to form. We also expect that the collision
zone for compound chondrule
formation must be quite thin, since the maximum time that the smaller secondary
can remain plastic is $\sim$ 20 s and the maximum collision velocity is
$\sim$ 0.1 km s$^{-1}$ ( a higher collision velocity would fragment the chondrules,
see Vedder and Gault (1974)). These numbers give a maximum length scale for
the thickness of the compound chondrule formation region of around 1 km.

Now, suppose we have a chondrule hover-particle in a region
of the flow where the ambient temperature of the gas flow
is much less than the solidus temperature of the particle.
If this solid hover-particle collides with smaller particles which are
entrained in the jet flow, and if the
relative speed of the two particles is greater than 0.1 km s$^{-1}$ then it is likely
that one or both of the particles will undergo some damage from the collision.
Depending on the relative speed of collision, this damage can range from
 slight chipping to complete fragmentation.

It is well known that many chondrules have undergone chipping and fragmentation. In the
Jet model, one can provide a cause for this damage: collisions between chondrules. One
can also give
 a site for where this damage will take place: in the jet flow at higher (and hence
colder) altitudes  than the compound chondrule formation region (see Fig.~\ref{fig:full_scheme}).
In this way, the jet model can provide a unified scheme
linking both compound chondrule formation
and chondrule fragmentation.

Of course, like all theoretical mutterings in this field, the above results and ideas
 should be (and
will be) treated with a reserved caution. A physical theory is of little use, unless
it can be used to predict as well as to explain. So far we have attempted the
latter, in the next section we will try the former. Let us throw caution to the winds
(no pun intended) and predict
the physical structure of triple compound chondrules.

\subsection{ The Weather Vane Effect}

\begin{figure}[ht]
\epsscale{0.3}
\begin{center}
\plotone{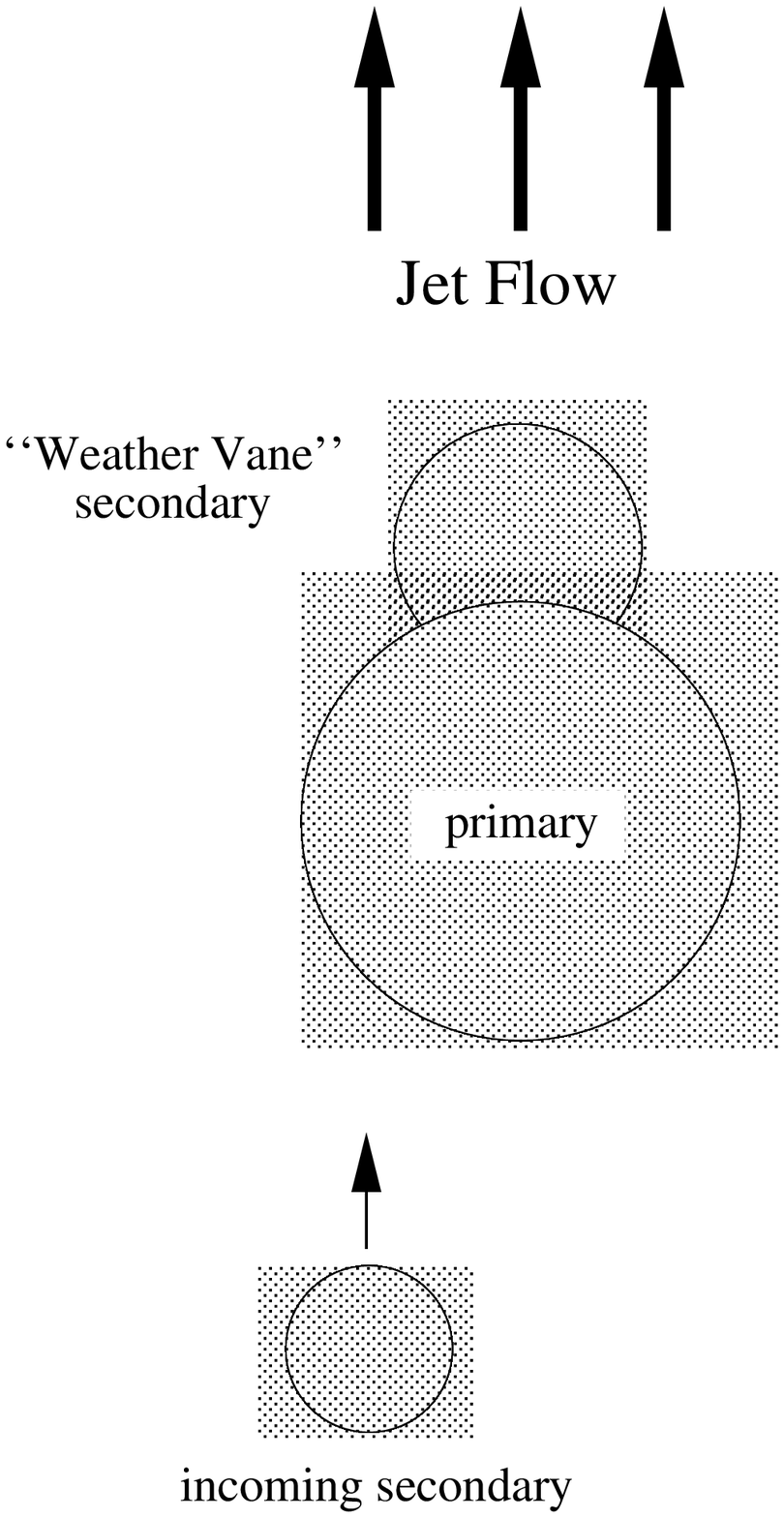}
\end{center}
\caption{
The ``Weathervane'' Effect. Secondary chondrules in a streaming gas
 flow, will tend to
orient the compound chondrule such that the secondary is 
pointing in the direction of the
flow. Any incoming chondrule will hit the rear end of 
the compound chondrule. Hence the
prediction that secondary chondrules in a triple
 compound chondrule will tend to avoid
each other.
}
\label{fig:weather_vane}
\end{figure}

If compound chondrules were formed in protostellar jets,
then gas drag would orient the
compound chondrules immersed in the jet flow. This is simply
because the spherical symmetry
of the simple chondrule is destroyed. A secondary chondrule will act
like the tail on a weather vane, orienting a binary compound chondrule such
that the smaller secondary will be on the ``downstream'' side of the primary.
A subsequent collision and fusion with another chondrule will produce a
triple compound chondrule where the secondary chondrules are separated by
a minimum avoidance angle. This situation is shown schematically in 
Fig~\ref{fig:weather_vane}.

This simple scenario does have some initial complications,
the major being that if the impact parameter (the offset distance between the
centres of the particles, see Fig~\ref{fig:weather_vane}) between
the colliding chondrules is greater than zero, then the binary compound
chondrule will be set spinning. This rotational motion will, however,
 be damped, because the  secondary
chondrule will be travelling half the time with the gas jet,
and half the time against the jet. If the secondary
chondrule has a cross-sectional area $A$,
then the rate of change of the rotation rate, $\omega$, is
given by
\beq
I\dot{\omega} \approx
\frac{C_{D} }{ 2}\rho_{g} a_{ps} (v_{gz}-a_{ps}\omega)^{2} \frac{A }{ 2}
- \frac{C_{D} }{ 2}\rho_{g} a_{ps} (v_{gz}+a_{ps}\omega)^{2}\frac{A }{ 2}
\ , 
\label{eq:3.3.1}
\eeq

where $I$ is the moment of inertia of the system, and $a_{ps}$ is the
distance between the primary and secondary chondrules.
 If we substitute the Epstein
solution for $C_{D}$ (see Eq. \ref{eq:3.1.4}) , we obtain
\beq
\dot{\omega} \approx
\frac{-8 }{ 3}\sqrt{\frac{8 k T_{g} }{ \pi m_{g}} }
\Bigl( \frac{\rho_{g} a_{ps}^{2} A }{ I } \Bigr) \omega  \ , 
\label{eq:3.3.2}
\eeq
which has the solution
\beq
\omega \approx \omega_{0}
\exp (- t / \tau_{sd} ) \ , 
\label{eq:3.3.3}
\eeq
where $\omega_{0}$ is the initial rate of rotation, and $\tau_{sd}$ - the
e-folding ``spin-down time''- has the form
\beq
\tau_{sd} = \frac{ 3I }{ 8 \rho_g A a^2_{ps} }
\sqrt{ \frac{\pi m_g  }{ 8 k T_g} } \ . 
\label{eq:3.3.4}
\eeq
To obtain an estimate for
 $\tau_{sd}$, we note that
\beq
I \sim (m_p + m_s) a^2_{ps} \ \lapp \ 2 m_p a^2_{ps} \ , 
\label{eq:3.3.5} 
\eeq
and
\beq
 A \sim \pi a^2_{s} \ , 
\label{eq:3.3.6} 
\eeq
where $m_p$ ($a_p$) and $m_s$ ($a_s$) are the masses (radii) of the primary and secondary
chondrules, respectively. Combining Eqns (\ref{eq:3.3.5}) and (\ref{eq:3.3.6}) with
the observation that $a_s \ \gapp \ 0.1 a_p$ and putting it 
all into Eq.(\ref{eq:3.3.4}) gives
\beq
 \tau_{sd} \ \lapp \ 6.5 \times 10^3 \
\frac{(\rho_p/ 3 {\ \rm g \ cm}^{-3}) (a_p / 0.1 {\ \rm cm}) }{
 (\rho_g/ 10^{-8} {\ \rm g \ cm}^{-3})  }
\sqrt{ {10^3 {\ \rm K} \over T_g} } \ s  \ . 
\label{eq:3.3.7}
\eeq
So for a range in gas densities of 10$^{-6}$ to 10$^{-10}$ g cm$^{-3}$
the spin-down timescales for a binary compound chondrule range from
around a minute to about a week.

\begin{figure}
\epsscale{0.4}
\begin{center}
\plotone{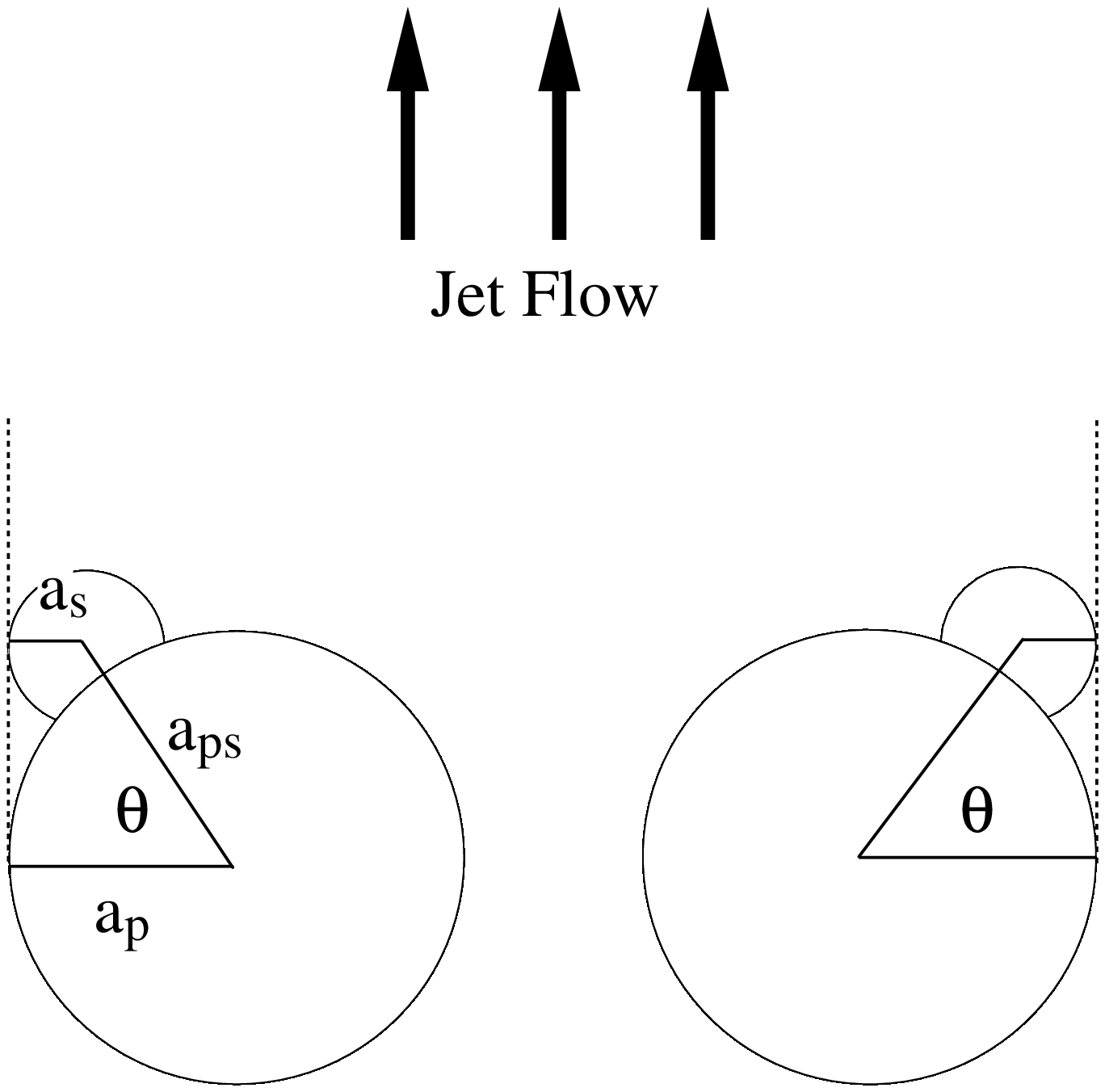}
\end{center}
\caption{
The extreme positions of the secondary chondrule while
 oscillating from side to side.
}
\label{fig:wobble}
\end{figure}

Once a binary compound chondrule has stopped spinning, then it will take
on a specific orientation, where the smaller secondary will be on the
downstream side of the primary chondrule. As shown in Fig~\ref{fig:wobble},
 the system
can still oscillate between two extreme positions, but there will always
be a section of the primary chondrule which will be shielded from colliding
with particles that are entrained in the gas flow.

From Fig~\ref{fig:wobble}, one can deduce a minimum avoidance angle $\theta $
that two secondary
chondrules on a triple compound chondrule should obey if the initial
binary compound chondrule has stopped spinning.
Given that the distance between the
centres of the primary and secondary chondrules is $a_{ps} =
a_{p}+fa_{s}$, where $f$
is the deformation factor of the secondary chondrule,
then
\beq
\cos(\theta) = \frac{a_{p}-a_{s} }{ a_{p}+fa_{s}}
         = \frac{1-a_{s}/a_{p} }{ 1+fa_{s}/a_{p}}. 
\label{eq:3.3.8}
\eeq

\begin{figure}
\epsscale{0.4}
\begin{center}
\plotone{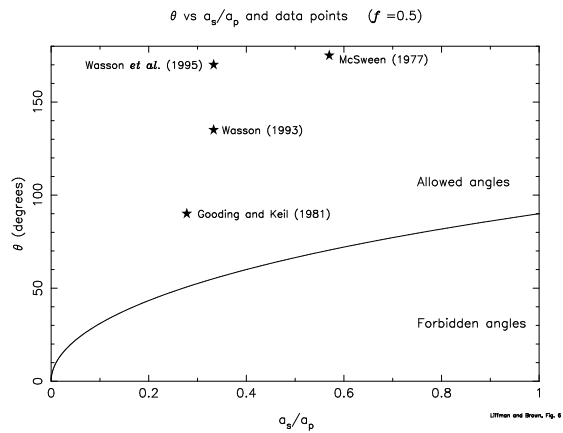}
\end{center}
\caption{
Comparison of theory vs observations for triple compound chondrules.
 We plot the angle between the centres of
secondary chondrules vs the radius ratio of the 
largest secondary to the primary.
The largest secondaries were used for $a_s$ values,
 as they would show `disallowed'
chondrules, if any were present. As can be seen, all the observed angles
lie in the `allowed' zone.
The line separating the `forbidden' and `allowed' zones is given by Eq.(\ref{eq:3.3.8}).
}
\label{fig:allowed_angles}
\end{figure}

A plot of a solution to this equation is shown in
Fig~\ref{fig:allowed_angles}. If the weather vane effect did not occur
then we would expect a uniform distribution of angles between secondaries
for each value of $a_{s}/a_{p}$. If the weather vane effect does occur,
then the distribution should be skewed towards the `allowed angles' shown
in Fig~\ref{fig:allowed_angles}. This is what we see in the
 few data we have been able to scour
from the literature.

Again, we should make clear that the weather vane effect will only be
observed if the timescale for a binary compound chondrule to stop spinning
is less than the chondrule-chondrule collisional timescales. It will also
only be observed in triple compound chondrules (i.e. one primary and two
adhering secondaries). Compound chondrules that have two or more adhering
secondaries will have unpredictable orientations and so an additional
secondary may land anywhere on such a compound chondrule.

Finally, we should point out that we have ignored gas flow fluctuations
that are parallel to the central plane of the accretion disk and
that the derivation of Eq.(\ref{eq:3.3.2}) assumes
 $v_{gz}$ remains constant during one revolution of the spinning
compound chondrule. This latter assumption is probably true just after the formation of the
compound chondrule, but may not be true as the compound chondrule stops spinning.

To see this, we note that a compound chondrule will stop spinning when it
 receives enough torque from the gas jet to remove all
the angular momentum and reverse the chondrule's initial spinning direction.
The length of time the secondary will be moving
against the gas jet is $\pi/\omega$. The torque is approximately
$C_{D}\rho_{g}v_{gz}^{2}Aa_{p}/2$.
So the minimum rotation rate $\omega_{min}$, is given by
\beq
I\omega_{min} \simeq
\frac{C_{D} }{ 2}\rho_{g}v_{gz}^{2}A a_{p} 
\frac{\pi }{ \omega_{min}}. 
\label{eq:3.3.9}
\eeq
Thus
\beq
 \omega_{min} \simeq
\biggl(\frac{\pi C_{D} \rho_{g} v_{gz}^{2}Aa_{p} }{ 2I} \biggl)^{1/2}. 
\label{eq:3.3.10}
\eeq
Using the Epstein drag law (Eq.(\ref{eq:3.1.4})) plus the approximate values for $I$ and $A$
 (Eqs (\ref{eq:3.3.5}) and (\ref{eq:3.3.6}), respectively). One can show that
\beq
 \omega_{min} \approx 0.015 \ {(a_s/0.1 \ {\rm cm}) \over (a_p/0.1 \ {\rm cm})^2 }
\biggl( \frac{(v_{gz} / 0.1 \ {\rm km s}^{-1} )(\rho_{g} / 10^{-11} \ {\rm g cm}^{-3} )
}{
(\rho_{p} / 3 \ {\rm g cm}^{-3} ) }
 \biggr)^{1/2}, \ {\rm rad \ Hz}   
\label{eq:3.3.11}  
\eeq
where we have assumed that $T_g = 1000$ K and that $m_g = m_H$.
So the period of rotation ($2 \pi / \omega$) of this slowly spinning compound chondrule is
approximately 400 seconds. In a 0.1 km s$^{-1}$ gas flow, 
this gives a scale length ($L$) of order
10 to 100 km.

So the relevant Reynolds number ($Re$) of the flow (see Liffman 1992) is given by
\beq
Re \approx 8.7 \times 10^3 \ {
(L/10^3 \ \hbox{km})(v_{gz} / 0.1 \ \hbox{km s}^{-1})(\rho_{g} / 10^{-11} \ {\rm g cm}^{-3} )
}{
\sqrt{( T_g / 10^3 {\ \rm K}) (m_H/m_g)} },  
\label{eq:3.3.12} 
\eeq
so, with $Re$ ranging from 100 to 1,000, $v_{gz}$ may undergo fluctuations which may modify
the spin-down time of Eq.(\ref{eq:3.3.4}).

\subsection{Chondrule Reheating}

If we combine the theory that we have acquired from \S 2 \& \S 3, we can present
a scenario for the phenomenon of chondrule reheating,
where high temperature rims are observed around chondrule cores
(Kring 1991). We suggest that reheating may be a consequence of
a chondrule overshooting its hover altitude and then oscillating, in a damped manner, around
the hover altitude. Each time the chondrule decreases its altitude, it will undergo an increase in
temperature. If the altitude decrease is large enough, the corresponding increase in
temperature will be sufficient to remelt the chondrule.

The actual mechanism whereby these high temperature rims are formed will not
be discussed in any depth here. We only suggest that as the chondrule
oscillates around its hover point, it accretes dust-grains/molten droplets
 and that this material
forms the foundation of the rim once the chondrule is reheated.

To understand how these oscillations arise, we note that
(see also \S 3.1) the equation of motion for a particle,
parallel to the $z$ axis,  is given by
\beq
 m_p\ddot{z} = \frac{C_D }{ 2} \rho_{g} (v_{gz}- \dot{z})^{2} \pi a_{p}^{2}
- \frac{ GMm_pz }{ (z^{2}+r^{2})^{3/2}} \ . 
\label{eq:3.4.1}
\eeq
As noted in \S 3.1, the drag coefficient ($C_D$) is likely to have the
Epstein form
\beq
 C_D \approx \frac{8 }{ 3v_{gz}} 
\sqrt{\frac{8kT_g }{ \pi m_g }} \ . 
\label{eq:3.4.2}
\eeq

Combining Eqs (\ref{eq:3.4.1}) and (\ref{eq:3.4.2}) and assuming $z \ll r$, we have:
\beq
m_p\ddot{z}
+ \frac{4 }{ 3}\Bigl( \frac{8\pi kT_{g} }{ m_{g}}\Bigr)^{1/2} \rho_{g}
\pi a_{p}^{2} \dot{z}
+\frac{GMm_pz }{ r^{3}}
\approx \frac{GMm_pz_{h} }{ r^{3}}, 
\label{eq:3.4.3}
\eeq
where $z_h$ is the hover height given by Eq.(\ref{eq:3.1.5}).

Let $x = z - z_h$, then Eq.(\ref{eq:3.4.3}) becomes
\beq
m_p \ddot{x} + c \dot{x} + k x = 0, 
\label{eq:3.4.4}
\eeq
 where
\beq
c = \frac{4 }{ 3}\pi a_{p}^{2} \rho_{g}\Bigl( \frac{8\pi kT_{g} }{ m_{g}}\Bigr)^{1/2},
\label{eq:3.4.5}
\eeq
and
\beq
k = \frac{GMm_p }{ r^{3}} . 
\label{eq:3.4.6} 
\eeq
 If we assume that $c$ and $k$ are approximately constant
 then $c^2 < 4km_p$ gives solutions of Eq.(\ref{eq:3.4.6}) that oscillate, with
 decreasing amplitude, as a function of time ($t$).
 If $c^2 = 4km_p$ or $c^2 > 4km_p$ then the solutions of Eq. (\ref{eq:3.4.6})
 are critically or
 strongly damped and the particle approaches its hover height without oscillation.

The case $c^2 = 4km_p$ translates into an equation which gives the critical
density for particle oscillation ($\rho_{gc}$):
\beq
\rho_{gc} = 2.7\times10^{-12} \ \frac{(a_{p}/0.1 {\ \rm cm} )
(\rho_{p}/1 {\ \rm gm} {\ \rm cm}^{-3})(M/{\rm M}_\odot)^{1/2}
(m_g/m_H)^{1/2}
}{
(r/0.1{\ \rm AU})^{3/2}(T/10^{3}{\ \rm K})^{1/2}
} \quad {\rm g cm}^{-3}. 
\label{eq:3.4.7}
\eeq
If $\rho_g < \rho_{gc} $ then $c^2 < 4km_p$ and the particle will oscillate
around the hover point $z_h$. It is our contention that it is this oscillation
process that causes chondrule reheating. Since the temperature of the
jet flow increases as $z$ decreases, thus as the particle travels to lower
altitudes its temperature must increase.

To be specific, when $\rho_g < \rho_{gc} $ then Eq.(\ref{eq:3.4.4}) admits the solution
\beq
 z = z_h + \frac{\dot{z}(z_h) }{ \mu}e^{-bt} \sin\mu t, 
\label{eq:3.4.8} 
\eeq
where $\dot{z}(z_h)$ is the $z$ speed of the chondrule when it first reaches $z = z_h$,
\beq 
\mu^2 = \frac{4 m_p k - c^2 }{ 4 m_p^2}, 
\label{eq:3.4.9} 
\eeq
and
\beq
 b = {c \over 2 m_p}. 
\label{eq:3.4.10}
\eeq
Thus the period of the oscillation is
\beq \frac{2 \pi }{ \mu} \ \gapp \ 11.6 \ {
(r/0.1{\ \rm AU})^{3/2} \over ({\rm M}_\odot)^{1/2}},
\quad {\rm days} 
\label{eq:3.4.11} 
\eeq
while the damping timescale is given by
\beq
\frac{1 }{ b} = 5 \ \frac{
(a_{p}/0.1 {\ \rm cm} )(\rho_{p}/1 {\ \rm gm} {\ \rm cm}^{-3})
(m_g/m_H)^{1/2}
}{
(\rho_g/10^{-12} {\ \rm gm} {\ \rm cm}^{-3})(T_g/10^{3}{\ \rm K})^{1/2} },
\quad {\rm days} 
\label{eq:3.4.12} 
\eeq
and the amplitude of the oscillation has the form
\beq \frac{\dot{z}(z_h) }{ \mu} =
1.6 \times 10^4 \ {
(\dot{z}(z_h)/ 0.1 {\ \rm km s}^{-1}) (r/0.1{\ \rm AU})^{3/2}
\over ({\rm M}_\odot)^{1/2}}. \quad {\rm km} 
\label{eq:3.4.13}
\eeq
Comparing Eqs (\ref{eq:3.4.13}) and (\ref{eq:3.1.6}) shows that if $\dot{z}(z_h)$ is
comparable to the gas flow speed, then the amplitude of the
oscillations can be a significant fraction of the hover height. The
corresponding temperature variations may also be significant.

Of course, from Eq.(\ref{eq:3.4.5}), $c$ is a function of $\rho_g$ and $T_g$
and cannot be considered a constant, so the above analysis should only
be treated as an informative approximation to the complete system.
A slightly more realistic analysis can be obtained by constructing a computer
simulation of the flow. We were able to accomplish this by assuming
a $z$ velocity for the outflow wind of the form
\beq
  \bar{v}_g(z) \approx 0.1 \ \biggl(\frac{z }{ z_h} \biggr)^m
\ \hbox{km s}^{-1},\qquad z \ \lapp \ z_h \ , \label{eq:3.4.14} 
\eeq
with $m = 0.1$, with the orbital (angular)
velocity of the gas assumed to be Keplerian.
The density structure of the gas flow could then be easily
 deduced from Eqs (\ref{eq:2.1}) and (\ref{eq:3.4.14}), and had the form
\beq
\bar{\rho}_g(z) \approx 4.5 \times 10^{-12}
\frac{(\dot{M}/10^{-8}{\ \rm M}_\odot/{\ \rm yr})
}{
( \bar{v}_g(z)/0.1 \ \hbox{km s}^{-1})(r/0.1{\ \rm AU})^2 }
\quad \hbox{g cm}^{-3},
\label{eq:3.4.15}
\eeq
while the temperature of the gas was given by
\beq
T_g(z) \approx 1200 \ \Bigl({z_h \over z}\Bigr)^{m(\gamma - 1)} \ {\rm K},
 \label{eq:3.4.16} 
\eeq
where we set $\gamma = 5/3 $.

As can be sen from the above formula, $T_g(z)$ will be greater than 1900 K
when $ z \ \lapp \ 0.001 z_h $. In such cases, we simply set $T_g(z)$ = 1900 K.
Finally, to compute the temperature of the particle, we did a full time
integration of Eq.(\ref{eq:3.2.1}).

\begin{figure}
\epsscale{0.4}
\begin{center}
    $ \begin{array}{c}
      \plotone{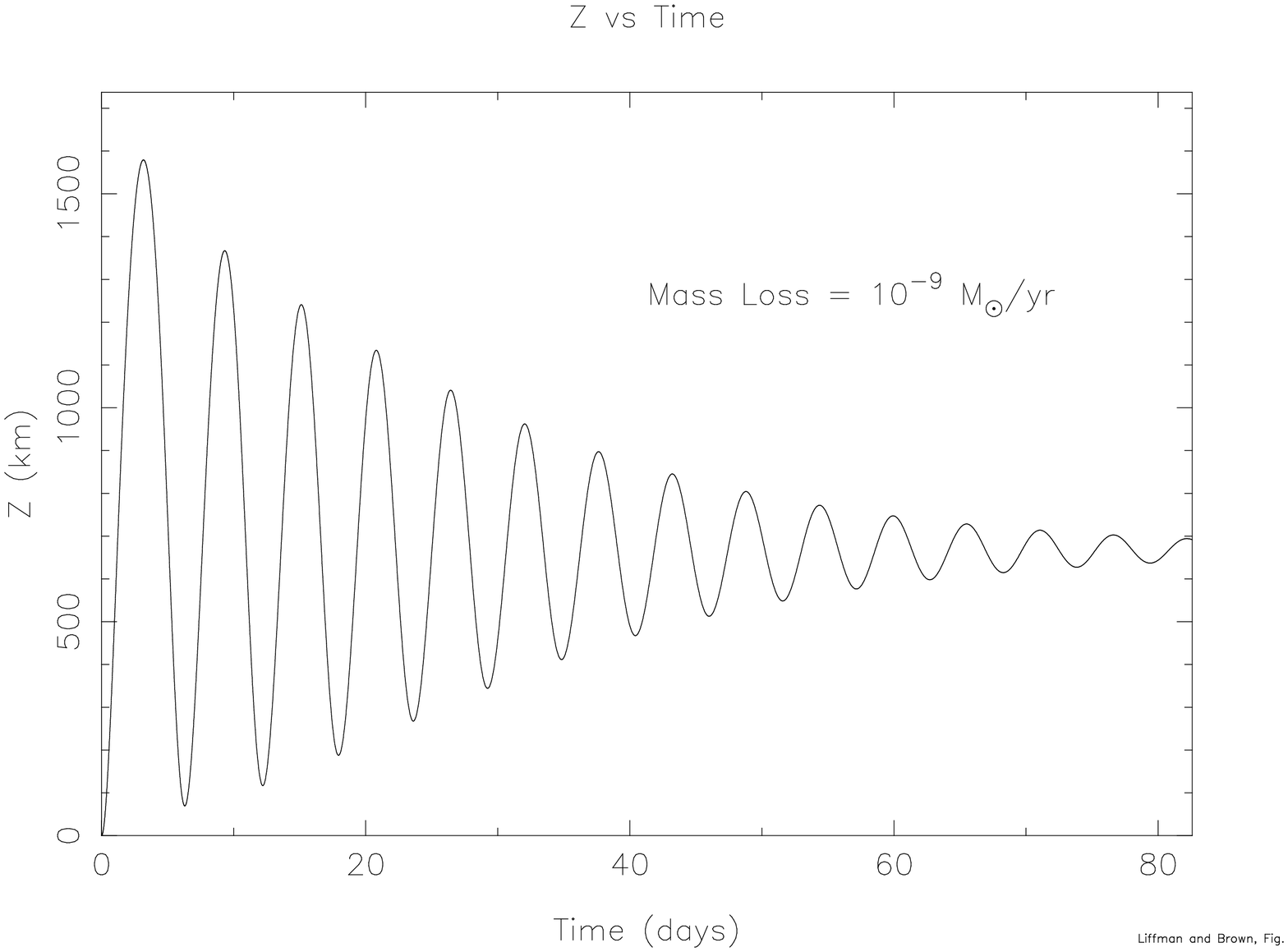} 
      \mathrm{(a)} 
      \plotone{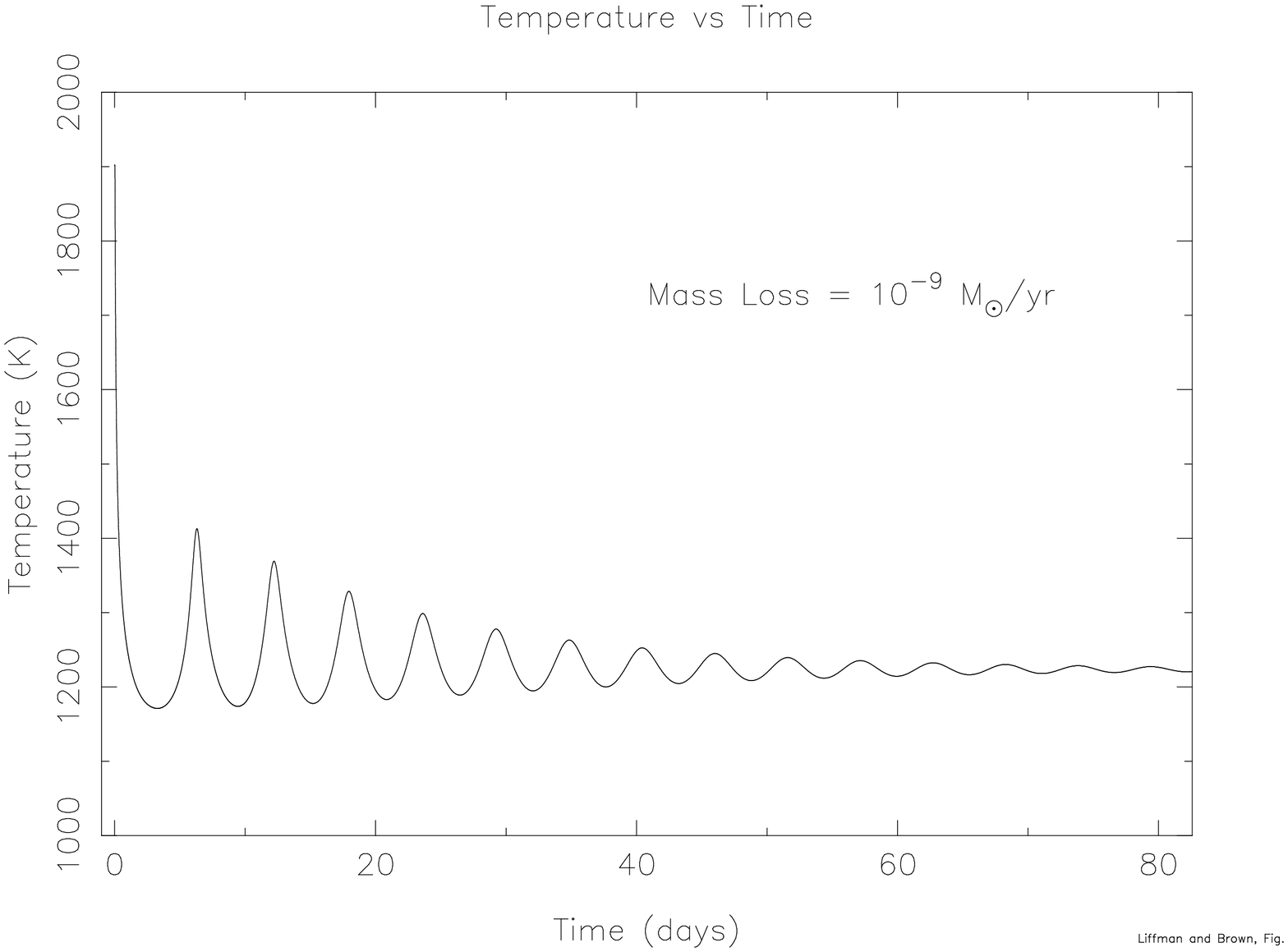} 
      \mathrm{(b)} 
      \end{array} $
\end{center}
\caption{(a) A chondrule-like particle is released into a
 protostellar jet flow at a distance
of 0.05 AU from a protostar.
The mass loss rate of the flow is set at
 $10^{-9}{\ \rm M}_\odot{\ \rm yr}^{-1}$.
Such a low mass loss rate ensures the gas density is lower than the critical
density given in Eq. (\ref{eq:3.4.7}). In such a circumstance 
the particle will oscillate,
in a damped manner, around the hover point (see Eq. (\ref{eq:3.4.8})).
 (b) The temperature of a particle as it undergoes the trajectory shown in
Fig.~\ref{fig:chondrule_reheat}a. }
\label{fig:chondrule_reheat}
\end{figure}

In Fig.~\ref{fig:chondrule_reheat}a
we show the motion of a chondrule-like particle ($a_p$ = 0.1 cm,
$\rho_p$ = 3.8 g cm$^{-3}$) released from very close to the midplane
($z$ = 100 cm) into a
jet flow with a total mass loss rate of $10^{-9}{\ \rm M}_\odot{\ \rm yr}^{-1}$.

The particle was released at a distance ($r$) of 0.05 AU from the center of a
solar mass protostar, where 0.05 AU is also the assumed radius of 
the protostellar jet.
The particle underwent a series of damped oscillations the period
 of which was slightly less than
6 days, while the damping timescale was approximately 30 days. The
corresponding temperature of the particle is shown in 
Fig.~\ref{fig:chondrule_reheat}b. The
particle temperature has a damped ``saw-tooth'' pattern. the maximum
temperature of the resulting temperature peaks being approximately 1420 K.
Of course higher temperature values can be obtained if the temperature
gradient is steeper than what we have assumed (\it e.g., \rm if $m > 0.1$).

\begin{figure}
\epsscale{0.4}
\begin{center}
     $\begin{array}{c}
      \plotone{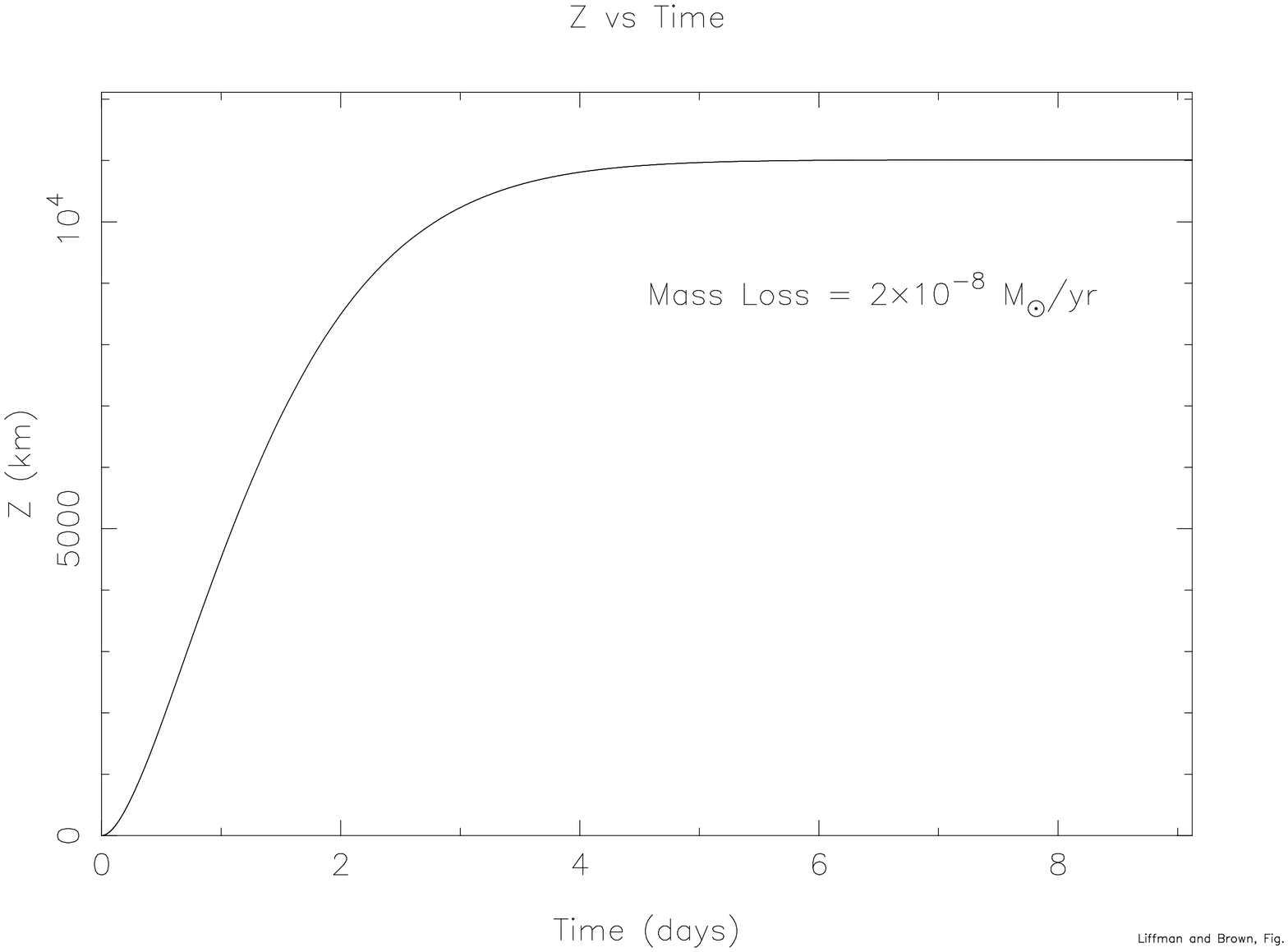}
      \mathrm{(a)}
      \plotone{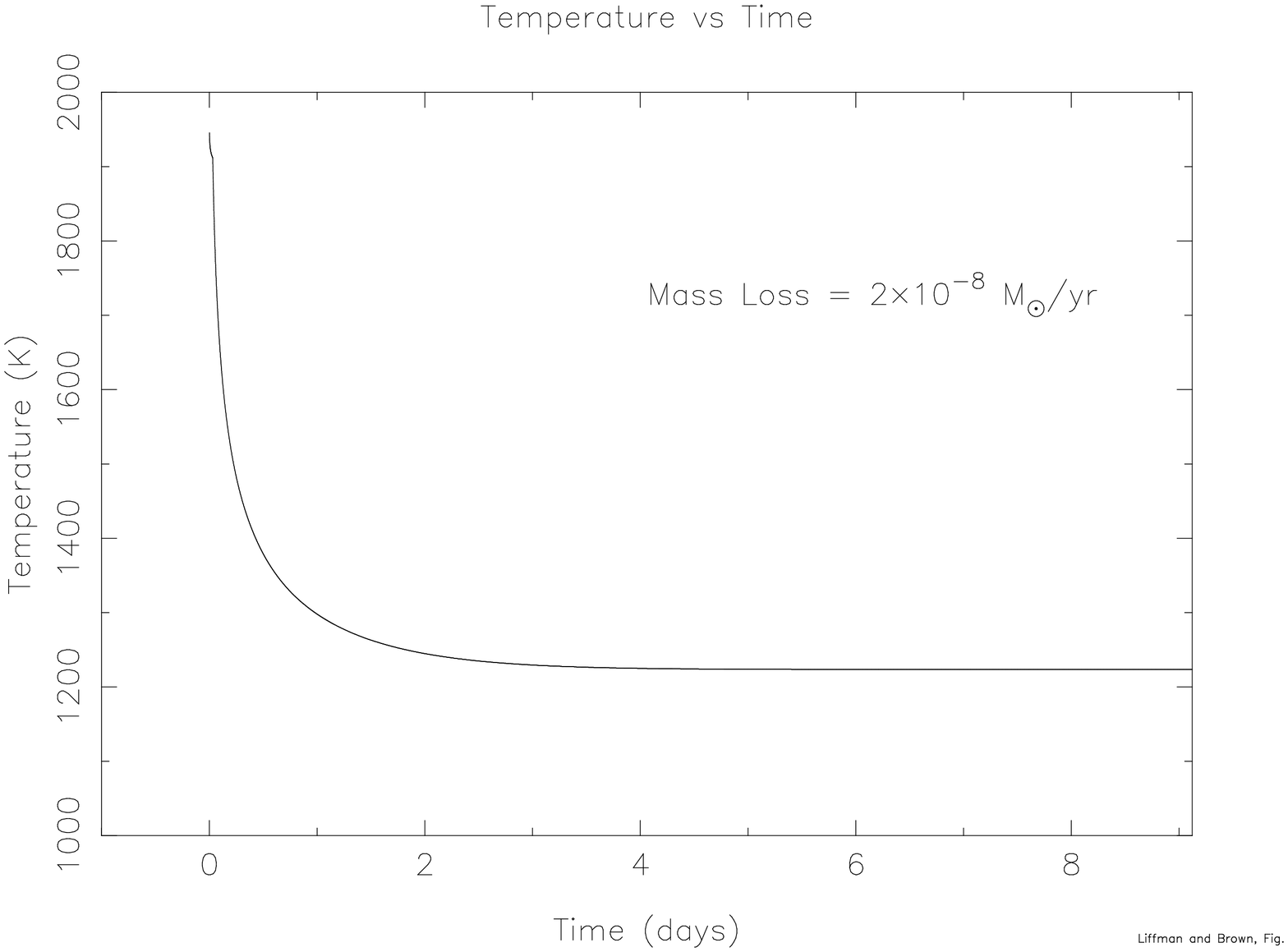}
      \mathrm{(b)}
      \end{array}$ 
\end{center}
\caption{(a) A chondrule-like particle is released into a 
protostellar jet flow at a distance
of 0.05 AU from a protostar.
The mass loss rate of the flow is set at $2 \times
10^{-8}{\ \rm M}_\odot{\ \rm yr}^{-1}$.
Such a low mass loss rate ensures the gas density is higher than the critical
density given in Eq. (\ref{eq:3.4.7}). In such a circumstance the particle will approach
the hover point in a critically damped manner (see Eq. (\ref{eq:3.4.17})).
 (b) The temperature of a particle as it undergoes the trajectory shown in
Fig~\ref{fig:chondrule_no_reheat}a. }
\label{fig:chondrule_no_reheat}
\end{figure}

While this oscillating motion is of interest, the particles can also
undergo a critically damped trajectory. In
particular, if  $\rho_g > \rho_{gc} $ then Eq. (\ref{eq:3.4.4}) gives the solution
\beq
z = z_h + \frac{\dot{z}(z_h) }{ 2\sqrt{-\mu^2}}
(e^{(-b+\sqrt{-\mu^2})t} - e^{(-b-\sqrt{-\mu^2})t}) , 
\label{eq:3.4.17} 
\eeq
The trajectory and temperature of such a particle are shown in
 Figs~\ref{fig:chondrule_no_reheat}a \& b,
where now the mass loss rate of jet is assumed to be
$2 \times 10^{-8}{\ \rm M}_\odot{\ \rm yr}^{-1}$.

From this simple model, we would suggest that chondrule reheating occurs
when the Jet flow is declining in mass flux, {\it i.e. \rm}, during the later stages of a
CTTS's evolution.

\section{THE CHONDRULE SIZE LIMIT}

Why do chondrules have an upper size limit? Possible solutions
such as size-limited precursor dust balls to aerodynamic sorting
have been suggested. In the Jet model, one has to eject particles
from the inner accretion disk and ensure that they arrive
in the outer parts of the accretion disk. Clearly, there must be
some size limit to this process and this size limit must
be dependent on the ``strength'' of the jet flow, which in turn
is dependent on the gravitational potential of the inner accretion
disk. To derive a quantitative relationship between chondrule and
accretion disk, we must first describe the behaviour of a droplet
in a flow.

Suppose we have a droplet of molten material that is subject  to a
streaming
gas flow. The surface drag of the flow will tend to
``rip''
the droplet apart, while the surface tension of the melt will try to
minimise the exposed surface area of the droplet and keep the
droplet
together. The balance between these two conflicting forces
produces a
stable
droplet of maximum radius $a_p$, the formula for which is
\beq
 a_p \approx { \gamma We_0  \over C_D\rho_gv_{gp}^2}, 
\label{eq:4.1}
\eeq
 where $\gamma$ is
the surface tension of the molten material, and $We_0$ is a
dimensionless factor
called the ``critical Weber number''. $We_0$ accounts for the
non-uniformity of the
gas drag pressure over the surface of the droplet. Typically, $We_0
\approx 10$ (Bronshten 1983).

The streaming gas flow will not only determine the stable size of
the
droplet, but
it will also subject the particle to a drag force given by
${C_D \over 2}\rho_gv_{gp}^2A_p$,  where $A_p$ is
the
cross sectional
area of the particle that is facing the gas flow.
\par
One can show (Liffman and Brown 1995) that the work done, $W$, by
the wind - flow
in ejecting a particle is
\beq
 W = {C_D^* \over 2}A_p<\rho_gv_{gp}^2>L,
\label{eq:4.2}
\eeq

where $C_D^*$ is a ``representative'' value of the drag coefficient
during the propulsion phase, $<\rho_gv_{gp}^2>$ is the mean value of
$\rho_gv_{gp}^2 $, and $L$
represents the actual length of the propulsion stage.
\par \vskip 4pt
For a particle to escape the disk, we require that
\beq
 <\rho_gv_{gp}^2> = {kGMm_p \over RC_D^*A_p L} , 
\label{eq:4.3}
\eeq
 where $k$ is $\geq$ 1, $G$ is the gravitational constant, $m_p$ is
the mass of
the particle, and $R$ is the initial radial (or semi-major axis)
distance of the particle from
the protostar. Substituting Eq.(\ref{eq:4.3}) into Eq.(\ref{eq:4.1}) gives
\beq
a_p \approx {We_oC_D^*\gamma RA_pL \over kC_D(0)GMm_p}, 
\label{eq:4.4}
\eeq
where $C_D(0)$ is the value of the gas drag coefficient at the position
where the particle is formed.
If we assume that our particle is
spherical (as are most unfragmented chondrules) and that our
propulsion distance $L$ is
proportional to the height of the disk at a distance $R$ away from
the protostar, i.e. $L = \lambda R^\beta$, then
\beq
a_p \approx \Bigl[ \frac{3 \lambda We_o  C_D^* \gamma R^{1+\beta} }{
4C_D(0)kGM\rho_p} \Bigr]^{1/2}. 
\label{eq:4.5}
\eeq

Now $C_D$, generally, decreases with increasing gas-flow speed. So, we should
expect that $C_D^*/C_D(0) \  \lapp \ 1$, since $C_D^*$ is an `average'
gas drag coefficient over the entire propulsion stage, while $C_D(0)$
samples the gas flow at the beginning of the propulsion stage,
where the gas flow is probably at its slowest.
 Setting $\lambda \approx 0.01, \beta \approx 1,$ (typical,
approximate values for the disk height), $ k \approx 1,
M \approx M_\odot, \hbox{and } We_o \approx 10 $, we obtain
\beq
 a_p \hbox{ (cm)} \ \lapp \ 0.4 \Bigl[ \frac{ \gamma }{ \rho_p }
\Bigr]^{1/2} R \hbox{ (AU)} . 
\label{eq:4.6}
\eeq

The values for the surface tension, $\gamma$, and the density,
$\rho_p$, are material
dependent. We are interested in Fe-Ni and
silicate chondrules, so we shall use
the surface tensions of meteoric iron and
stone, which are $\gamma_{iron} =
1,200$ and $\gamma_{stone} = 360$ gs$^{-2}$ ( Allen et al. 1965)
with corresponding mass densities of
$\rho_{iron} = 7.8$ and $\rho_{stone} = 3.4$ gcm$^{-3}$.
Substituting these values into Eq.(\ref{eq:4.6})
gives us the approximate radius of iron and silicate droplets as a
function of distance from
the protostar:
\beq
 r_{iron} \hbox{(cm)} \  \lapp \ 5R  \hbox{ (AU)}
\hbox{
and } r_{stone}
\hbox{ (cm)} \ \lapp \ 4R \hbox{ (AU)} .  
\label{eq:4.7} 
\eeq

If chondrules were formed from this ablative process, then to obtain
 the observed maximum chondrule sizes we require that R $\lapp$ 0.1
AU (and $k \geq 1$ ).
This is consistent with the conclusion that
protostellar jets are formed in an accretion disk at a distance of
0.05 to 0.1 AU from a
solar-type protostar
 (Camenzind 1990, Hartmann 1992).

Although this result is encouraging, there is a contradiction between
the  ideas presented in \S 3 and \S 4. In \S 3, we require the chondrules
to  remain stationary in the flow so to allow the formation of compound chondrules,
while in this section we have shown that the chondrule size limit is determined by
the ``power''  of the jet flow  to eject particles from the inner accretion disk.
This contradiction would  appear to severely limit, perhaps destroy, our theory.
After all,  how can one require particles to be stationary relative to the accretion
disk and also expect them to be ejected at speeds $\geq$ the escape
speed of the protostellar system?

The only way out of this contradiction is for the jet flow to be highly
variable in both density and velocity. Compound chondrules would presumably
form when the flow
is relatively quiescent, while chondrule ejection would perhaps
occur when there is a major
increase in the density and/or velocity of the flow.
Observations do suggest that jet flows are highly variable in their behaviour
(Edwards \it et al. \rm 1993).
For example, Mundt (1984)
obtained observational evidence of variations in young stellar winds
that vary on time scales of months.
Despite  this observational support, the resolution of this
problem must await a coherent theory of protostellar jet formation,
which in turn must await high resolution observations of protostellar jets.

Besides determining a maximum size limit for chondrules,
the ejection of chondrules by a protostellar jet also, indirectly, causes
chondrules and chondrule fragments to be size sorted.

\section{ SIZE SORTING AND DISK ACCRETION}

It has been known for many years (Dodd 1976) that chondrules, both
silicate and metal, are size-sorted in meteorites, i.e., particles in a
particular chondrite satisfy the relation (Skinner and Leenhouts 1993)
\beq
 \rho_g a_p \approx {\rm constant} \ .
\label{eq:5.1}
\eeq
 It has recently
become apparent that size-sorting also applies to fragments of
chondrules (Skinner and Leenhouts 1991). This shows that chondrites
act as ``size-bins''. Chondrules and their fragments were formed, mixed
and later sorted by size within the chondrite forming regions of the
solar nebula.

 The usual explanation for this phenomenon
(e.g. Dodd 1976 and references therein), is some form of aerodynamic
drag effect, where larger and/or denser particles can travel further
into a resistive medium than can smaller and/or less dense particles.
We too shall use this idea, for it arises naturally from the Jet model.

To see this, suppose that a particle is ejected from the inner accretion
disk at speeds close to or exceeding the escape
velocity. Suppose further that the angular
momentum of the particle is high enough to eject the particle from
the gas flow and allow it to move across the face of the accretion
disk. If the particle is not subject to gas drag, from the upper-atmosphere
of the accretion disk,
 it will simply move out of the
system and into interstellar space.
If the particle is subject to gas drag then, given sufficient drag,
the particle will be recaptured by the protostellar system and
may fall into the outer parts of the accretion disk.

A stream of such particles will be size-sorted. Since the smaller, less
dense particles will fall close to the protostar, while the larger,
more dense particles will fall further away from the protostar.
Thus, we should expect that meteorites from the outer
parts of the solar system will contain larger particles than
meteorites from the inner solar system, and indeed, chondrules from the carbonaceous
chondrites are (usually) larger than chondrules from ordinary chondrites.

There is an additional consequence of this model which
has to do with angular momentum transfer. A particle in Keplerian orbit
around a star has an angular momentum that is proportional to the
square root of the distance between the particle and the star. Thus,
a particle that is ejected from the inner accretion disk, and subsequently
stopped by gas drag,
will simply fall back into its original orbit unless angular momentum
is transferred from the disk to the particle. If gas drag allows
the disk to transfer angular momentum to the particle, then the particle
can fall to the outer parts of the disk.

 Clearly, if the disk gas gives up angular momentum, then it must
move in towards the protostar. Since protostellar jets are
fueled by disk accretion, one obtains the schematic picture of a jet
flinging out material into the disk such that the disk will accrete
onto the protostar and refuel the jet. Heuristically, one can
think of chondrules and other associated Jet ejecta as delayed protostellar
Jet fuel.

To turn this qualitative speculation into quantitative speculation,
we have to model the particle ejection process. Unfortunately, this is
a difficult thing to do, since there is no consensus on
how protostellar jets work. So, we make the following assumptions:

(1) We suppose that our chondrule  particles are, initially, in a circular Keplerian orbit
of radius $R \  ( \leq 0.1$ AU) from the protostar.

(2) We assume that the protostellar jet
 gives the particles an initial ``boost'' velocity $\dot z(0)$ that is comparable to the
protostar's escape velocity at that point. Our tentative justification
for this assumption is that protostellar jets are observed to have
speeds comparable to the maximum escape velocity of a protostellar system.
So, we presume that particles initially entrained in such a flow may also obtain
 similar speeds.

(3) Finally we assume that the particles have, initially,
no radial velocity, and that the self-gravity of the disk is negligible
compared to the gravity of the protostar. The former of these two assumptions
comes from the observation that jet flows tend to be perpendicular to
their respective accretion disks. Protostellar jets do have a
 nonzero radial velocity, but we have, for simplicity, ignored this component.
The validity or otherwise
of these ideas is discussed at some length in Liffman and Brown (1995).

Given these assumptions, the equations of motion for a particle become
\beq
\ddot r  = \frac{h^2 }{ r^3} - \frac{ GMr }{ \bigl[ r^2 + z^2
\bigr]^{3/2} }\ , 
\label{eq:5.2}
\eeq
\beq
v_{\theta} =  r\dot \theta = {h \over r}\ , 
\label{eq:5.3}
\eeq
and
\beq
\ddot z = - \frac{ GMz }{ \bigl[ r^2 + z^2 \bigr]^{3/2} }\ , 
\label{eq:5.4}
\eeq
where $h$ is the specific angular momentum of the particle and
has the
value
\beq
 h = \sqrt{GMR} \ . 
\label{eq:5.5}
\eeq
The value of $h$ is a
constant, since it is assumed that there are no external torques,
parallel to the $z$ axis,
acting on the system.

\begin{figure}
\epsscale{0.7}
\begin{center}
\plotone{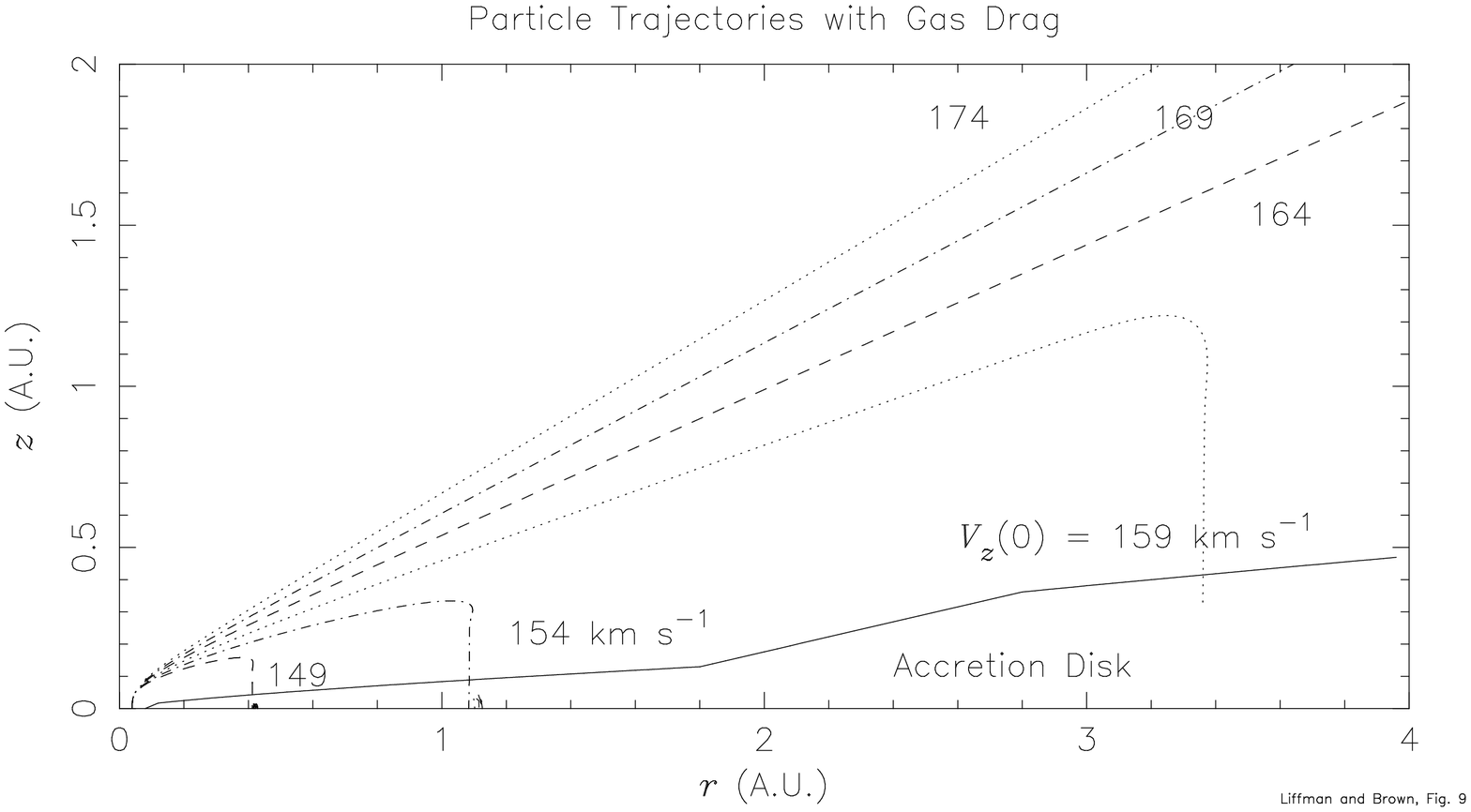}
\end{center}
\caption{
Simulation data showing chondrule-like particles travelling above the disk and
falling back into it. A solar-mass protostar is located at $r = 0, z = 0$. Surrounding
the protostar is an accretion disk, the scale height of which is shown in profile.
Chondrule-like particles (radius = 0.1 cm, density = 3.5 g cm$^{-3}$) are given
a velocity boost in the $z$ direction, the magnitudes of which (in km s$^{-1}$)
are shown next to the
trajectories. Particles that are subject to sufficiently high gas drag are later recaptured.} 
\label{fig:trajectories}
\end{figure}

These equations are not difficult to model numerically and can be
solved by
standard techniques. Although a brief analysis of the above
equations shows that
if the particles are travelling at speed greater than the
escape velocity of the system, then the initially vertical
path of the projectile will quickly turn into a ``horizontal''
path across the face of the accretion disk.
A computer simulation of this phenomenon is given in 
Fig.~\ref{fig:trajectories}
( see also Liffman and Brown (1995)).

In Fig.~\ref{fig:trajectories},
 particles are ejected at $r$ = 0.04 AU with different
vertical velocities ranging
from 149 to 174 km s$^{-1}$. The particles are assumed to move
out of the gas
outflow at $r$ = 0.1 AU, whereupon they encounter the gas halo of
the accretion
 disk and their subsequent motion is governed by gas drag  plus
the gravitational force from the protostar.

As the particles move through the halo gas of the accretion disk,
they will acquire, by gas drag, angular momentum from the disk.
 If we assume that the centrifugal
force of the halo gas
balances  the radial component of the protostar's gravity, then the
angular speed of the halo
gas, $v_{\theta, gas}$, will be given by
\beq
v_{\theta, gas} = \frac{r\sqrt{GM} }{ (r^2 + z^2)^{3/4}} . 
\label{eq:5.6}
\eeq
 Once the particle has come to rest, relative to the halo gas, it
will have the angular velocity given by Eq.(\ref{eq:5.6}) and a specific
angular momentum, $h$, given
by $h = rv_{\theta, gas}$. As can be deduced from Eq.(\ref{eq:5.2}), such a
specific angular  momentum
implies that $\ddot r = 0$, and the only force acting on the particle
will be the $z$
component of the gravitational force, which will point towards the
accretion disk. As the
particle moves towards the accretion disk, the gas density and
angular
velocity of the gas will increase, thereby keeping $\ddot r \approx 0$.
As can be seen from Fig.~\ref{fig:trajectories}, the subsequent path of the particle is roughly
parallel to the $z$ axis.

The paths of the captured particles, shown in Fig.~\ref{fig:trajectories},
 can be approximated
to that shown in Fig. 10, where the ascending path length, \it l \rm, is
given by $ l = K \chi$, with $K$ being a number in the range of $4 \pm 1$,
(Liffman and Brown 1995)
and $\chi$ is the ``stopping distance'' as defined by the equation,
\beq
\chi = \frac{4a_p\rho_p }{ 3\rho_g} \ . 
\label{eq5.7}
\eeq
As a justification for the above equation,
we note that a macroscopic particle will come to rest when it has
encountered a total gas mass approximately equal to its own mass.
 The ``stopping distance'',
 so defined by this prescription, is easily shown to be the $\chi$
length scale as given in the above equation.
So, for a constant gas density, the larger or more dense a particle
is, the further it will be able to travel before it comes to rest.




\begin{figure}
\epsscale{0.3}
\begin{center}
\plotone{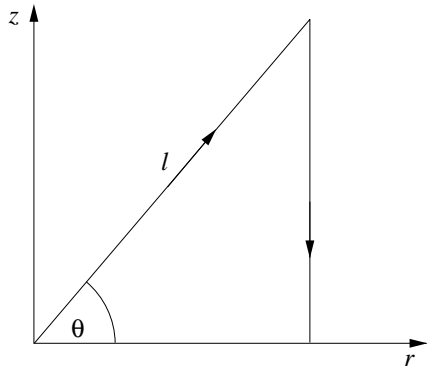}
\end{center}
\caption{
An approximation of chondrule trajectories as a triangular path.
} 
\label{fig:schematic_path}
\end{figure}

 The range of  a recaptured projectile is simply given by the formula
\beq
 r \approx l \cos(\theta) \, , 
\label{eq:5.8}
\eeq
where $\theta$ is the angle between the ascending path of the particle
and the midplane of the accretion disk.

The simplicity of Eq.(\ref{eq:5.8}) suggests that aerodynamic size sorting may occur
to particles that are recaptured by the protostellar system. To see this, we
 suppose that we have two particles with different mass densities: $\rho_1$ and
$\rho_2$. Suppose further that these two particles fall back to the accretion
disk at the same distance from the protostar. Then we can write the radius
ratio of particle 1 to particle 2 as
\beq
\frac{a_1 }{ a_2} \approx \frac{ \rho_2 }{ \rho_1 } Q(r) \, , 
\label{eq:5.9}
\eeq
where
\beq
Q(r) = \frac{ K_2 }{ K_1 }\frac{ \rho_{g1} }{ \rho_{g2}}
 \frac{ \cos( \theta_2 )}{ \cos( \theta_1) }\, . 
\label{eq:5.10}
\eeq
The factor $Q(r)$ is dependent on the initial ejection
speeds of the particles (through $K$ and $ \theta$), the initial distance of the particles
 from the protostar ( again $K$ and $ \theta$)
and the scale height of the accretion disk (through $\rho_g$).
If the particles are created at about the same distance from the protostar
then we would expect that $K, \theta$ and $\rho_g$
would be similar for both particles, since the average flight path of both particles
would be about the same. This would imply that $Q(r) \approx 1$.
So, from the collection of particles
that fall to the accretion disk at $r$, the denser
particles should have smaller radii.

As has been discussed, size
sorting is density dependent in that dense Fe-Ni chondrules are always smaller than
less dense silicate chondrules. The mass densities of the two types of
chondrules are (Skinner and Leenhouts 1993) $\rho_{Si} \approx$ 3.8 g c$m^{-3}$,
 and $\rho_{Fe} \approx$ 7.8 g c$m^{-3}$,
 which implies that $ \rho_{Si} / \rho_{Fe} \approx 0.5 $,
and so if Eq.(\ref{eq:5.2}) is applicable to chondrules, we should expect that
\beq
 \frac{a_{Fe} }{ a_{Si}} \approx 0.5 Q(r) \, . 
\label{eq:5.11}
\eeq

Using published data for the
meteorite types H, L, LL, (Dodd, 1976) and CR (Skinner and Leenhouts 1993),
we can compute the average Fe to Si size ratio for all these different
types of meteorites (excluding Bjurb$\ddot{\hbox{o}}$le)
\beq
\biggl< \frac{<a_{Fe}> }{ <a_{Si}>} \biggr>_{H,L,LL,CR} = 0.52 \pm 0.16  \ . 
\label{eq:5.12}
\eeq
The corresponding approximate mean $Q$ values is
\beq
<Q_{H,L,LL,CR}> =1.04 \pm 0.32 \ . 
\label{eq:5.13} 
\eeq

Of course, the agreement between the above $Q$ value and our
theoretical model should be treated with caution, since we have only presented the
bare beginnings of a quantitative model. Nonetheless, it does illustrate the
potential of the ``Jet'' model to explain the phenomenon of size-sorting.

Finally, we return to our discussion of angular momentum transfer and disk accretion.
Let us consider a ring of material in an accretion disk at a distance $r$ from a
protostar. The angular momentum of this material is
\beq
 L(r) = m(r) \sqrt{GMr} \ , 
\label{eq:5.14}
\eeq
where $m(r)$ is the mass of the ring of material at $r$. Now suppose that material
 falls onto the accretion disk, and this infalling material has essentially zero
angular momentum. In such a case, the angular momentum of the ring is conserved, and
the mass of the ring becomes a function of time, i.e., $m \equiv m(r,t)$.
For such a case, we can differentiate Eq. (\ref{eq:5.14}) to obtain
\beq
 \frac{dr }{ dt} = - \Bigl( \frac{2r }{ m} \Bigr) \frac{dm }{ dt} \ . 
\label{eq:5.15} 
\eeq
or
\beq
 r(t) = r(0) \Bigl( \frac{m(0) }{ m(t)} \Bigr)^2 \ . 
\label{eq:5.16}
\eeq
Thus, if infalling material doubles the mass of the ring, it will move
from its initial position $r(0)$ to $r(0)/4$. It is via this mechanism of mass
and angular momentum transfer that disk accretion may be, in part, mediated. Indeed, because
it is the halo gas of the accretion disk that will be transferring most of the
angular momentum to the ``Jet projectiles'', it is possible that the upper layers
of the accretion disk are the ones that undergo most of the accretion, leaving
the midplane relatively untouched.

  This type of process may explain an implicit contradiction between
observations and meteoritics. Observations suggest that protostars keep on
accreting material from their disk for periods of up to 10$^7$ years
(Cabrit \it et al. \rm 1990).
 Radiometric
data from the decay of $^{129}$I and $^{26}$Mg suggest that meteorites accreted
material for periods of order 10$^6$ - 10$^7$ years. How could the meteoritic
material have been preserved if a major portion of the solar nebula was accreted
onto the protosun?  The answer, we suggest, is that disk accretion was, in part,
altitude dependent. Material flung from the protosolar jet mediated the angular momentum
transfer and one component of this mass transfer was the chondrule.

\section{CHONDRULE-MATRIX COMPLEMENTARITY}

In this paper, we claim
that chondrules have the physical characteristics
expected of ablation droplets that have been formed and ejected by a protosolar jet,
 and then recaptured by the solar nebula through the action of
 gas drag.

Such a model, however, is immediately confronted with the complementary composition of
matrix and non-matrix material in meteorites. For example,
  Wood (1985) discusses the case of Murchison, where the matrix has an Fe/Si ratio of
 1.23, while the non-matrix material
(chondrules, CAIs, isolated crystals) is $\sim$ 0.2. These two dissimilar components
combine to give an Fe/Si ratio of 0.81, which is close to the solar value of 0.9.
This is unlikely to be accidental, and is clear evidence for the local formation of
chondrules, and the refractory component of chondritic material.

Before we throw out our wind-transport model, however, one should note
 that chondrules, at least in
this model, form in a relatively small region of the solar nebula, i.e.,
in or near the boundary layer of the protostellar system. If the chondrules
have a low Fe/Si value then the surrounding material will, by mass balance, have a high Fe/Si
ratio. If this latter  material comes in a non-gaseous form
( e.g 10 $\mu$m dust, CAI \&c) then it  too will be
ejected with the chondrules. One will obtain the desired mass balance
if all this material lands back into the solar nebula, in a uniform manner,
 and over a long period of time.

For such a model to work, we require a fairly large component of matrix
material to be made from dust that has been recycled through the protostellar
jet. There are at least two consequences if this idea is correct.
First, this recycled dust would have to be more refractory than CI  material,
since the jet formation region of the inner solar nebula would have been far warmer
than the regions where most chondrites were formed.
Second, if matrix material were formed from dust that had been lofted into the
upper atmosphere of the solar nebula then this may be an observable phenomenon.

To understand this latter point, we need an estimate for the amount of dust
that should be resident in the upper atmosphere of the solar nebula
at any particular time.

Protostellar jets
have average mass loss rates of order
$10^{-8} \ {\rm M}_\odot{\rm yr}^{-1}$ and since
the mass of dust to gas in the Interstellar Medium (ISM) is 1/100, this
implies that $\lapp 10^{-10} \ {\rm M}_\odot{\rm yr}^{-1}$ of dust
is blown out by the
protostellar jet. So, an upper limit for the amount of dust that is lobbed into
the upper atmosphere of an accretion disk in a Classical T Tauri Star (CTTS) per year is
$10^{-10} \ {\rm M}_\odot$.

The dust settling timescale ($\tau_{settle}$) can be deduced from Eq. (\ref{eq:3.1.1})
with $v_{gz} = 0$, {\it i.e. \rm},
\beq
m_p \ddot z = - \frac{C_D }{ 2} \rho_g \dot z^2 A_p - \frac{GMm_pz
}{ [r^2 + z^2]^{3/2}}
 \ . \label{eq:6.1} 
\eeq
The motion of the dust is subsonic, so the
drag coefficient takes
the Epstein form for $C_D$ (Eq. \ref{eq:3.1.4}). To
compute the  velocity
of the particle, we note that $\ddot z \approx 0$, which implies
\beq
 \dot z \approx \frac{- \omega_K^2(r,z)\rho_p a_p }{ \rho_g \bar v } \ z
\ , 
\label{eq:6.2}
\eeq
where $\omega_K(r,z) = \sqrt{GM}/[r^2+z^2]^{3/4}$ is the
Keplerian angular velocity at the point $(r,z)$, and $\bar v = \sqrt{8kT_g/\pi m_g}$
is the mean Maxwellian speed of the gas particles.

 If $z \ll r$ then
$\omega_K(r,z) \approx \omega_K(r) = \sqrt{GM/r^3}$ and we can
compute the settling
time:
\beq
 \tau_{settle} \approx 8,000 \ \frac{\bigl(\rho_g/ 10^{-11} \hbox{ g cm}^{-3} \bigr)
\bigl( \bar v/1 \hbox{ km s}^{-1} \bigr) \bigl( r/1 \hbox{ AU} \bigr)^3   }{
\bigl(a_p/1 \mu \hbox{m} \bigr) \bigl(\rho_p/1 \hbox{ g cm}^{-3} \bigr) }
\ \hbox{ yr} \ . 
\label{eq:6.3}
\eeq
Thus, we have characteristic dust-settling timescales in the range
$10^3 - 10^4$ years, which means that up to $10^{-7} - 10^{-6}$ M$_\odot$ of dust
will be in the upper atmosphere of a CTTS accretion disk at any one time.

As discussed in Natta (1993), $10^{-7}$ M$_\odot$ of high-altitude dust may produce
the observed ``flat-temperature distributions'' in CTTSs. These temperature
distributions are a surprisingly common phenomenon in CTTSs. They arise
when the temperature
of the disk does not decrease as rapidly with distance from the protostar
as one would predict from
standard accretion disk theory. Natta suggested that a spherical halo of dust
around a CTTS would reflect light from the protostar and into the accretion
disk, thereby increasing the temperature of the outer disk.

Of course, the idea that protostellar jets can loft dust into the
outer parts of the surrounding accretion disk requires a quantitative
 investigation to determine whether dust grains can be ejected
from the jet flow. There
will be a size limit where particles smaller than a certain size
will simply be entrained in the jet flow and ejected from the system.
 We simply note, that we
require dust ejection similar to that shown in \S 5
 to account for chondrule-matrix complementarity.

\section{CONCLUSIONS}

Chondritic meteorites are typically an agglomeration of igneous rocks, i.e.,
chondrules and refractory inclusions (formation temperatures 1500-2000K)
, surrounded by sedimentary material that,
in some cases, has never experienced temperatures greater than 500K.
This unusual structure has prompted theorists to develop
heating mechanisms (e.g. lightning) that can provide brief, intense
impulses of energy in the otherwise
cold outer regions of the solar nebula. These energy
impulses are presumed to have melted small
dust aggregates into chondrules, which were then incorporated  into larger dust
aggregates that eventually formed meteorites.

Such energy impulse theories are not required in the `Jet' model of chondrule formation.
Chondrules are formed in the hot inner regions of the accretion disk adjacent to the
protostar. There is no difficulty in obtaining the required temperatures, because
in or near this region the accretion disk dumps around half of its gravitational
energy. The protostellar jet ejects the chondrules from the hot inner disk
and gas drag brings these particles back to the cooler outer regions, where they could
be incorporated into growing aggregates of cool nebular material.
The ``plum pudding'' structure of chondritic meteorites is a natural consequence of
this model.

One of the many problems in chondrule formation is the deduced low
 cooling rate ($\sim$ 1 - 1000 K/hour) for these particles. This cooling
rate is many orders of magnitude smaller than that expected for a particle
radiating directly into space. In our model, chondrules are formed in the
optically thick regions of the inner accretion disk. They will, therefore,
be in thermodynamic equilibrium with the gas and their temperature variations will be
damped.

Of course if the chondrules were
to remain in this environment, their temperatures would not decrease.
Chondrules, however, are produced by the ablative interaction between
a streaming gas flow (perhaps produced from the merging of magnetic field lines)
and molten material. Particles, that are small enough, will be swept up with
the flow. As the gas flow moves away from the midplane of the
inner accretion disk, it will expand and therefore probably cool.
Particles that are entrained in this gas flow will also cool at the same rate.

Particles that are moving with the gas flow may be ejected from the
accretion disk, but it is possible that a particle may simply hover
at some distance away from the midplane of the disk. If the gas density in the flow is
below a critical gas density ($\sim 10^{-11}$ g cm$^{-3}$) then the particles
will undergo damped oscillations around their hover points. It is due to these
oscillations, we suggest, that chondrules can undergo reheating. Smaller particles
of the same mass density will still move past the hovering particles, and
so may collide with these particles.

Such a scenario allows for the formation of adhering compound chondrules.
These chondrules pose a major problem for chondrule formation
theories as it is nearly  always the smaller chondrules that were plastic at the
time of collision. Smaller particles will lose heat more readily than
larger particles, since they have a larger surface to area ratio.
If chondrules were formed in a single flash heating event, we should expect
the smaller particles to become solid before the larger particles.
That we actually see the opposite behaviour
(i.e., the large particles were solid, while the small particles were
semi-molten), strongly suggests that single flash-heating models
require some modification.

In the Jet model, adhering compound chondrules are formed when
small chondrules collide  with larger chondrules that are stationary in the flow.
 The smaller particles that are moving with the flow
will be warmer than the larger hovering particles, because the latter particles have
had time to equilibrate with the local temperature of the gas. If the temperature
gradient in the flow is sufficiently steep ($\sim$ 10 to 100 K/km) at this collision point,
compound chondrules with the observed structure will be formed.

Once a compound chondrule has been formed, and has stopped spinning,
the smaller secondary chondrule
will orient the entire compound chondrule, just like a weathervane,
such that the secondary chondrule is pointing in the direction of the flow.
Any incoming secondary  chondrule will hit the rear end of the primary chondrule.
The resulting triple compound chondrule, will have two secondaries chondrules that
will tend to ``avoid'' each other. Of course, this prediction implicitly assumes
a faily steady gas flow, an assumption that may be incorrect. However,
if this prediction is found to be valid, it
would be major piece of evidence in favour of the Jet model, as it is difficult to
produce such an effect with other chondrule formation theories.

The fragmentation of chondrules is a simple extension of compound chondrule
formation, where instead of one particle being plastic at the time of collision,
we now have two solid particles colliding at higher velocities. We
expect the fragmentation zone to be `above' the compound-chondrule formation zone,
since the gas flow will be cooler and the flow speed higher as one moves
further away from the midplane of the disk.

The chondrule size range is yet another aspect of chondrule formation which
has not been satisfactorily explained. In the Jet model, the sizes of the
ejected droplets are determined by the balance between gas drag and the surface tension.
A jet flow with high energy density will tend to make molten droplets smaller, while
molten materials with higher surface tensions will tend to form larger droplets.
Observations suggest that protostellar jets are produced within 0.1 AU of solar-mass
 protostars. The minimum energy density of a wind that can eject Si and Fe droplets from such close
proximity to the protostar is such that the radii of these droplets is $\leq $ 1 cm.
Protostellar winds that form larger droplets, will have a lower energy
density and will not be able to eject them from the jet
formation region and we will not see them in meteorites.

This ejection mechanism may also explain the complementary chemical structure of
chondrules and their surrounding matrix. For this to occur, we require that refractory
dust as well as chondrules are ejected by the jet flow so that mass balance will
be obtained once the dust settles back to the solar nebula. This leads to a simple
calculation, which suggests that up to $10^{-7} - 10^{-6}$ M$_\odot$ of dust
will be in the upper atmosphere of a CTTS accretion disk at any one time.
These figures appear to be consistent with those deduced from observation (Natta 1993).

The transfer of dust and chondrules from the inner to outer portions of the
 accretion disk, necessarily requires the removal of angular momentum from the
 upper atmosphere of the accretion disk. This will increase the viscosity of the
disk and hasten
the stratified infall of disk material, \it i.e., \rm the upper atmosphere of the disk
will accrete more readily than the central plane of the disk.

Finally, aerodynamic size sorting of the ejected particles is a natural consequence of
the Jet model. Ejected particles that are captured by the accretion disk must suffer
aerodynamic size sorting, since the particles will be subject to gas drag. Particles
that are not subject to gas drag will simply leave the protostellar system, since
their initial velocities were higher than the escape velocity of the system.

It is for all these reasons that we consider chondrules to be ablation droplets
formed by a protosolar jet in the first 10$^6 - 10^7$ years of the solar system.

\begin{figure}
\epsscale{0.6}
\begin{center}
\plotone{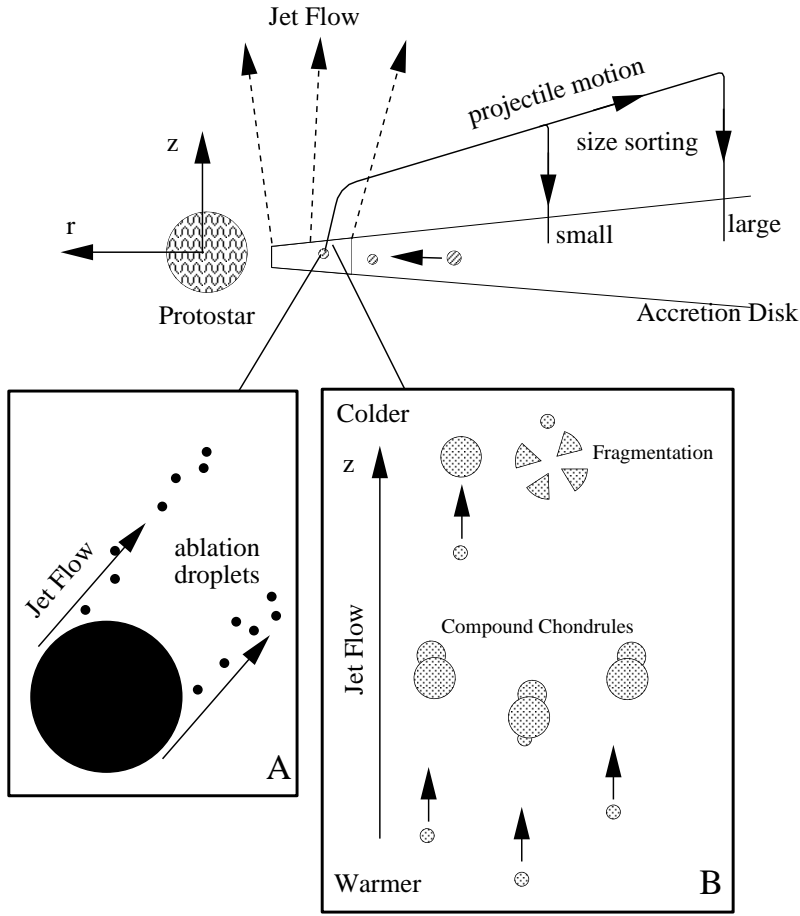}
\end{center}
\caption{
A pictorial summary of the ideas presented in this paper.
An accretion disk surrounds a protostar. At a distance within
0.1 AU of the protostar, the disk produces a protostellar jet
(only one side of which is shown). Infalling km-sized bodies
move into this hot ``jet zone'' and are subsequently ablated by the
jet wind. The resulting droplets, if small enough, move with the flow and
away from the ablating body. It is our claim that these ablation
melt-droplets eventually become chondrules.
At higher altitudes, within the accretion disk, some of the larger particles
may hover in the gas flow. Smaller particles will still move with the flow,
however, and may collide with the hover particles. These chondrule-chondrule
collisions produce compound
chondrules and chondrule fragments. The fragmentation of chondrules occurs
at higher altitudes relative to the compound-chondrule formation
zone, because the flow
cools as it increases in altitude and the two colliding chondrules will be
solid at the time of collision.
If the protostellar jet is sufficiently powerful, it will eject particles
from the inner accretion disk, and if the orbital angular momentum of the
ejected particles is large enough, the particles will move out of the
jet flow and travel across the face of the accretion disk. With
sufficient gas drag from the upper atmosphere of the accretion disk,
these (now size-sorted) particles will be brought down
 to the outer accretion disk, where
they will be incorporated into planetesimals.
}
\label{fig:full_scheme}
\end{figure}

\acknowledgments
 {\bf Acknowledgements}

The first author wishes to extend his thanks to the Program Committee
of the ``Chondrules and the Protoplanetary Disk'' Conference and
the LPI for their generous travel grant. Dr Rachel Webster is similarly
thanked for the matching grant given to the first author by the School of
Physics at the University of Melbourne.

\vskip 20pt

\bf
\centerline{REFERENCES}
\rm

\vskip 10pt

\noindent

\baselineskip 16pt
Allen, J. H., B. S. Baldwin B.S. Jr., and James N.A. (1965)
\it Effect on Meteor Flight of Cooling by Radiation and Ablation\rm.
\bf NASA TN D-2872\rm.
\vskip 10pt

Beckwith, S.V.W., and Sargent A.I. (1993)
The occurrence and properties of disks around young stars.
 In \it Protostars and Planets III \rm,
( E. H. Levy, and J. I. Lunine, Eds), pp. 521-541.
 University of Arizona Press.
\vskip 10pt

Bicknell G. V. 1992. Mechanisms for the production of bipolar flows
and jets in star formation regions. \it Aust. J. Phys. \rm \bf 45, \rm
513-529.
\vskip 10pt

Bronshten V. A. (1983) \it Physics of Meteoric Phenomena. \rm Reidel,
Dordrecht.
\vskip 10pt

Brownlee D. E., Bates B., and Beauchamp R. H. 1983
Meteor ablation spheres as chondrule analogs. In
\it Chondrules and Their Origins \rm (E. A. King, Ed.), pp. 10-25.
Lunar and Planetary Institute, Houston.
\vskip 10pt

Cabrit S., Edwards S., Strom S.E., and Strom K.M. 1990
Forbidden-line emission and infrared excess in T Tauri stars:
Evidence for accretion driven mass loss?
  \it Astrophys. J. \rm \bf 354, \rm 687-700.
\vskip 10pt

Camenzind M. (1990) Magnetized disk-winds and the origin of
bipolar outflows. In \it Accretion and Winds\rm, (G. Klare, Ed.),
  Rev. Modern Astron. 3, 234-265. Springer-Verlag.
\vskip 10pt

Cassen P. (1993) Why convective heat transport in the solar nebula was
inefficient. \it Lunar Planet. Sci. Conf. \bf 24, \rm 261-262.
\vskip 10pt

Dodd, R. T. (1976) Accretion of the ordinary chondrites.
 \it Earth Planet. Sci. Lett.  \bf 30, \rm 281-291.
\vskip 10pt

Dodd, R. T., and L. S. Teleky (1967) Preferred orientation of olivine
crystals in porphyritic chondrules. \it Icarus \bf 6, \rm 407-416.
\vskip 10pt

Edwards S., Ray T., and Mundt R. (1993)
Energetic mass outflows from young stars.
 In \it Protostars and Planets III \rm,
( E. H. Levy, and J. I. Lunine, Eds), pp. 567-602.
 University of Arizona Press.
\vskip 10pt

Epstein P. S. (1924) On the resistance experiment by spheres in their
motion through gases. \it Phys. Rev. \bf 2, \rm 710-733.
\vskip 10pt

Frank J., King A., and Raine D. (1992)
\it Accretion Power in Astrophysics. \rm Cambridge University Press,
Cambridge. 294 pp.

Freeman J. 1977 The magnetic field in the Solar nebula
\it Proc. Lunar Sci. Conf. 8th \rm 751-755.

\vskip 10pt

Fukui Y., Iwata T., Mizuno A., Bally J. and Lane A. P. (1993)
Molecular Outflows.  In \it Protostars and Planets III \rm,
( E. H. Levy, and J. I. Lunine, Eds), pp. 603-639.
 University of Arizona Press.
\vskip 10pt

Gooding, J., and K. Keil (1981) Relative abundances of chondrule
primary textual types in ordinary chondrites and their bearing on
conditions of chondrule formation. \it Meteoritics \bf 16, \rm 17-43.
\vskip 10pt

Hartmann L. (1992)
Winds from protostellar accretion disks.
In \it Nonisotropic and Variable Outflows from Stars \rm,
(L. Drissen, C. Leitherer, and A. Nota, Eds),
A.S.P. Conference Series, \bf 22, \rm
pp. 27-36.
Astronomical Society of the Pacific.
\par
\vskip 10pt

Hewins R.H. (1988) Experimental studies of chondrules.
 In \it
Meteorites and the Early Solar System \rm (eds J. F. Kerridge and M. S.
Matthews), pp. 660-679. University. of Arizona Press.
\vskip 10pt

K$\ddot{\hbox{o}}$nigl, A., and Ruden S. P. 1993. Origin of outflows and winds.
 In \it Protostars and Planets III \rm,
( E. H. Levy, and J. I. Lunine, Eds), pp. 641-687.
 Univ of Arizona Press, Tucson.
\par
\vskip 10pt

Kring D.A. (1991) High temperature rims around chondrules in primitive
chondrites: evidence for fluctuating conditions in the solar nebula.
\it Earth and Planetary Science Letters \bf 105, \rm 65-80.
\par
\vskip 10pt

Liffman, K. (1992) The formation of chondrules by ablation. \it Icarus \rm \bf
100, \rm 608-619.
\par
\vskip 10pt

Liffman, K. and Brown M. (1995) The motion and size sorting of particles
ejected from a protostellar accretion disk. \it Icarus \rm in press.
\par

Lovelace, R. V. E., H. L. Berk, AND J. Contopoulos (1991)
Magnetically driven jets. \it Astrophys. J. \bf 379, \rm 696-705.
\par
\vskip 10pt

McSween H. (1977) Chemical and petrographic constraints on the
origin of chondrules and inclusions in carbonaceous chondrites.
\it Geochim. Cosmochim. Acta. \bf 41, \rm 1843-1860.
\par
\vskip 10pt

Mundt R. (1984) Mass loss in T Tauri stars: Observational
studies of the cool parts of their stellar winds and expanding
shells. \it Ap.J. \bf 280, \rm 749-770.
\par
\vskip 10pt

Natta A. (1993) The Temperature Profile of T Tauri Disks.
\it Ap. J. \bf 412. \rm 761-770
\par
\vskip 10pt

Priest E. R. (1994) Magnetohydrodynamics.
In \it Plasma Astrophysics \rm.
(A. O. Benz and T. J. - L. Courvoisier, Eds), pp. 1-109.
Spinger-Verlag.
\par
\vskip 10pt

Shu F., Najita J., Galli D., Ostriker E., and Lizano S. (1993)
The collapse of clouds and the formation and evolution of stars and disks.
 In \it Protostars and Planets III \rm,
( E. H. Levy, and J. I. Lunine, Eds), pp. 3-45.
 University of Arizona Press.
\par

Shu F., Najita J., Ostriker E., Wilken F., Ruden S., and Lizano S. (1994)
Magnetocentrifugally driven flows from young stars and disks. I. A
generalized model.
\it Astrophys. J. \bf 429, \rm 781-796.
\vskip 10pt

Skinner, W.R., and J. M. leenhouts (1991)
Implications of chondrule sorting and low matrix contents of type 3
ordinary chondrites. \it Meteoritics \rm  \bf 26, \rm 396.
\par
\vskip 10pt

Skinner, W.R., and J. M. leenhouts (1993)
Size distributions and aerodynamic equivalence of metal chondrules and
silicate chondrules in Acfer 059. \it  Lunar Planet. Sci. Conf.  \bf 24, \rm
 1315-1316.
\par
\vskip 10pt

Snell R.L., Loren R.B. and Plambeck R.L. (1980)
Observations of CO in L1551: evidence for stellar wind driven shocks.
\it Astrophys. J. Lett. \bf 239, \rm 17-22.
\par
\vskip 10pt

Swindle T.D., Caffee M.W., Hohenberg C.M., Lindstrom M.M. and
Taylor G.J. (1991) Iodine-xenon studies of petrographically and chemically
characterized Chainpur chondrules. \it Ceochim. Cosmochim. Acta \bf 55, \rm
861-880.
\par
\vskip 10pt

Swindle T.D., and Podosek F.A. (1988) Iodine-Xenon dating. In \it
Meteorites and the Early Solar System \rm (eds J. F. Kerridge and M. S.
Matthews), pp. 1127-1146. University. of Arizona Press.
\par
\vskip 10pt

Vedder J.F., and Gault D.E. (1974) A chondrule: evidence of
energetic impact unlikely. \it Science \bf 185, \rm 378-379.
\par
\vskip 10pt

Wasson J.T. (1993) Constraints on chondrule origins. \it Meteoritics \bf
28, \rm 14-28.

\vskip 10pt

Wasson J. T., Krot A. T., Lee M. S., and Rubin A. E. (1995)  Compound Chondrules.
Submitted to \it Meteoritics \rm.

\vskip 10pt

Weidenschilling S.J. (1977) Aerodynamics of solid bodies in the solar nebula.
\it Mon. Not. Roy. Astron. Soc. \bf 180, \rm 57-70.

\vskip 10pt

Weidenschilling S.J. (1988) Formation processes and timescales for meteorite
parent bodies. In \it Meteorites and the Early Solar System \rm
(eds J.F. Kerridge and M.S. Matthews), pp. 348-371. University. of Arizona Press.

\vskip 10pt

Whipple F.L. (1973) Radial pressure in the solar nebula as affecting the motions of
planetesimals. \bf  NASA-SP-319, \rm 355-361.

\vskip 10pt

Wood, J. A. 1985. Meteoritic constraints on processes in the
solar nebula.
 In \it Protostars and Planets II \rm,
( D. C. Black, and M. S. Matthews, Eds), pp. 687-702.
 Univ of Arizona Press, Tucson.

\end{document}